\documentclass{emulateapj}
\def\stacksymbols #1#2#3#4{\def\theguybelow{#2}
        \def\verticalposition{\lower#3pt}
        \def\spacingwithinsymbol{\baselineskip0pt\lineskip#4pt}
        \mathrel{\mathpalette\intermediary#1}}
\def\intermediary #1#2{\verticalposition\vbox{\spacingwithinsymbol
        \everycr={}\tabskip0pt
        \halign{$\mathsurround0pt#1\hfil##\hfil$\crcr#2\crcr
                \theguybelow\crcr}}}
\def\lta{\stacksymbols{<}{\sim}{2.5}{.2}}
\def\gta{\stacksymbols{>}{\sim}{3}{.5}}

\shorttitle{DYNAMICS INSIDE CYGNUS A}
\shortauthors{MATHEWS \& GUO}

\begin{document}

\title{Dynamics inside the radio and X-ray cluster cavities 
of Cygnus A and similar FRII sources}


\author{William G. Mathews\altaffilmark{1} 
and Fulai Guo\altaffilmark{1}}

\altaffiltext{1}{University of California Observatories/Lick
Observatory,
Department of Astronomy and Astrophysics,
University of California, Santa Cruz, CA 95064
mathews@ucolick.org}

\begin{abstract}
We describe approximate axisymmetric computations of the dynamical
evolution of material inside radio lobes and X-ray cluster gas
cavities in Fanaroff-Riley II sources such as Cygnus A.  All energy is
delivered by a jet to the lobe/cavity via a moving hotspot where jet
energy dissipates in a reverse shock.  Our calculations describe the
evolution of hot plasma, cosmic rays (CRs) and toroidal magnetic
fields flowing from the hotspot into the cavity.  Many
observed features are explained.  Gas, CRs and field flow back
along the cavity surface in a ``boundary backflow'' consistent with
detailed FRII observations.  Computed ages of backflowing CRs are
consistent with observed radio-synchrotron age variations only if
shear instabilities in the boundary backflow are damped and we assume
this is done with viscosity of unknown origin.  A faint
thermal jet along the symmetry axis may be responsible
for redirecting the Cygnus A non-thermal jet.  Magnetic fields
estimated from synchrotron self-Compton (SSC) X-radiation observed
near the hotspot evolve into radio lobe fields.  Computed profiles of
radio synchrotron lobe emission perpendicular to the jet are
dramatically limb-brightened in excellent agreement with FRII
observations although computed lobe fields exceed those observed.
Strong winds flowing from hotspots naturally create kpc-sized spatial
offsets between hotspot inverse Compton (IC-CMB) X-ray
emission and radio synchrotron emission that peaks 1-2 kpc ahead where
the field increases due to wind compression.  In our computed version
of Cygnus A, nonthermal X-ray emission increases from the hotspot
(some IC-CMB, mostly SSC) toward the offset radio
synchrotron peak (mostly SSC).
\end{abstract}

\vskip.1in

\section{Introduction}

Iconic radio and X-ray images of Cygnus A (Figure 1) 
attest to the colossal energy 
attributed to cluster-centered massive black holes.
Apparently in response to mass accretion,
the massive black hole in Cygnus A 
ejects opposing non-thermal jets that form 
radio lobes and X-ray cavities.
The jets in Cygnus A 
penetrate out through the cluster gas, 
driving strong bow-shocks that enclose the jet 
and its cavity like a cocoon.
Since the velocity of the jet greatly 
exceeds that of the expanding shock, 
a second more powerful (reverse) shock 
must appear near the apexes of the cocoon where 
the energy of the jet 
is delivered to kpc-sized post-shock hotspots. 
The velocity of the hotspot is relatively modest, 
similar to that of the 
bow shock, but gas and relativistic particles flow 
through the hotspot with much higher velocities. 
Matter flowing from the high pressure hotspots 
inflates the entire radio lobe, displacing the cluster gas
as it forms an X-ray cavity. 
Most, or perhaps all, of the contents of the radio lobes 
-- relativistic particles, magnetic
field and plasma -- originated in the bright hotspots.
As hotspots move out into the cluster gas, 
the energetic CRs produced there 
flow back (in a ``backflow'') toward the cluster center. 

Images as in Figure 1 have inspired 
many theoretical studies of FRII jets 
and their cocoons 
(e.g. Blandford \& Rees 1974; Scheuer 1974; 
Kaiser \& Alexander 1997; 
Clarke, Harris, \& Carilli 1997;
Carvalho \& O'Dea 2002;
Carvalho et al. 2005; 
Krause, 2005;  
Saxton et al. 2002;
O'Neill \& Jones 2010;
Huarte-Espinosa, Krause, \& Alexander 2011).  
Here we describe approximate calculations 
emphasizing the dynamical evolution of material 
inside the radio lobes.


The approximate evolution and current morphology
of Cygnus A can be produced by cosmic rays flowing away 
from the hotspot as it moves out into the cluster gas
(Mathews \& Guo 2010, hereafter MG10). 
The powerful jet compresses the hotspot in the reverse shock 
at its inner surface, 
but most of the energy inside hotspots 
is contained in cosmic rays 
transported from the jet and/or accelerated in the strong 
reverse shock.
Jet transport of CRs may be more likely since 
normal diffusive shock acceleration
is suppressed by the magnetic field orientation 
perpendicular to the jet 
(Sironi \& Spitkovsky 2009) as observed by Carilli et al. (1999).
In the discussion below, as in MG10, 
we regard the hotspot as the primary energy source in the cocoon. 
With this assumption we avoid direct computation 
of the jet itself which occupies a very small volume (Fig. 1) 
and which is much more difficult to observe and interpret 
than the bright hotspots, visible in both radio and X-rays (Fig. 1). 
Multiple pairs of 
discrete hotspots (as in Figure 1) are common 
in FRII sources (Black et al. 1992;
Hardcastle, Croston, \& Kraft 2007), 
indicating that the direction of opposing jets 
changes rather abruptly from time to time. 
These changes can create a new hotspot before the previously 
activated  
hotspot has decayed and are sufficiently abrupt 
not to produce visible cometary 
smears where moving jets impact on the relatively denser 
outer wall of the radio cavity. 

Here we describe calculations similar to those 
in MG10 but with an emphasis on the detailed 
flow of post-hotspot gas and cosmic rays inside the radio cavity. 
In particular, 
we address two of the difficulties encountered in the MG10 
computations: (1) large scale irregularities in the 
radio lobe boundaries (Figs. 3, 9 \& 11 in MG10) that 
have no observed counterparts  
and (2) chaotic, high velocity plasma flows inside the radio 
lobes that are inconsistent with the regular, 
radially ordered age-related variation of 
synchrotron spectra observed in Cygnus A (Fig. 1) and 
in FRII sources in general 
(Alexander \& Leahy, 1987). 
The orderly variation of observed radio spectra along the radio lobes  
reveals an evolutionary aging as 
cosmic ray (CR) electrons lose energy by 
synchrotron emission. 
Associated with this is a well-ordered internal flow pattern.
Evidently 
CRs and the magnetic field necessary for synchrotron emission 
both advect along with low density gas  
as it backflows from the hotspot toward the cluster center.
As a result,  
the oldest CR electrons are found closer to the center of Cygnus A, 
furthest from their hotspot origin.
The monotonic radial variation of synchrotron ages 
(e.g. Alexander \& Leahy 1987; Machalski et al. 2007) in FRII sources 
indicates that their advection 
must be spatially smooth, uninterrupted by turbulence or 
large scale non-laminar flows inside the radio cavity.

Clearly, the vortical irregularities in the lobe boundary 
and the chaotic flows inside the radio lobe that appear 
in previous FRII computations must not occur.
The surface vortices appear to be 
non-linear Kelvin-Helmholtz (KH) instabilities driven by shear 
between rapidly backflowing post-hotspot material 
and adjacent gas on both sides.
KH irregularities also drive 
vortical and other large scale flows deep inside 
the radio cavities that 
spatially mix synchrotron emitting CR electrons 
of all ages, upsetting the 
highly stratified age variation observed. 
Disordered internal lobe velocities computed by MG10  
are sufficiently large, $\gta 500$ km s$^{-1}$, 
to distort the fragile
arrangement of radio-synchrotron ages 
created during the Cygnus A lifetime, about $10^7$ yrs.
In MG10 we recognized that some 
damping mechanism must be invoked to reduce or remove 
these shear-generated disturbances.
Our impression is that 
similar unobserved internal lobe velocities are 
common in all previous computational
studies of FRII evolution. 

KH instabilities can be stabilized by strong magnetic 
fields along the lobe or by viscous damping.
In view of the dynamical weakness of 
observed magnetic fields in Cygnus A, 
we explore here the possibility that the apparent absence 
of KH vortical activity is due to viscosity. 
We do not claim to understand the physical nature 
of viscosity in a collisionless 
relativistic fluid mixed with very low density 
plasma and weak fields.
However,
a similar viscous damping has a remarkably beneficial effect in
removing unobserved surface irregularities in the gamma ray
image of the Fermi bubbles in the Milky Way
(Guo \& Mathews 2012; Guo et al. 2012).
Viscosity in hot cluster gas has also been considered 
by Reynolds et al. (2005), Roediger \& Bruggen (2008) and
Jones (2008).

Our objective is to explore dynamical and physical features  
in Cygnus A created by hot gas, CRs, and magnetic fields 
using 2D axisymmetric computations and a variety of 
additional simplifying assumptions. 
While we adopt many parameters consistent with observations 
of Cygnus A and its surrounding cluster gas,
we do not adjust parameters to achieve the best possible 
match to Cygnus A observations -- 
our results are intended to apply to FRII sources in general.
We adopt a distance to Cygnus A of $\sim230$ Mpc so that 
1$\arcsec$ corresponds to 1 kpc.

\section{Computational Procedure}

As discussed in MG10, we consider the self-consistent dynamics 
of a two-component fluid: 
relativistic cosmic rays (CRs) and hot gas. 
The pressures of these two fluids are related to the 
energy densities by $P_c = (\gamma_c - 1)e_c$ 
and $P = (\gamma - 1)e$ respectively where 
$\gamma_c = 4/3$ and $\gamma = 5/3$. 
CR pressure gradients communicate momentum to the gas 
by means of small magnetic fields frozen into the gas, 
assuming that Alfv\'en speeds are generally small 
compared to typical gas velocities. 
The magnetic energy density $u_B = B^2/8\pi$ 
inferred from radio and X-ray 
observations of the Cygnus A radio lobes 
is smaller than $e_c$ by factors of 10-600 
(Hardcastle \& Croston 2010; Yaji et al. 2010). 
Observed fields in the radio lobe are 
$15-20\mu$G (Yaji et al. 2010). 
Even the much larger magnetic fields 
observed in Cygnus A hotspots, $\sim 200\mu$G, 
indicate that $u_B$ is several times smaller than $e_c$ 
(Stawarz et al. 2007).
Consequently, for the approximate computations discussed 
here we ignore the Lorentz force 
${\bf j \times B}$ on the gas and 
regard the magnetic field as passively moving with the 
hot gas velocity.

The equations we consider are:
\begin{equation}
{ \partial \rho \over \partial t}
+ {\bf \nabla}\cdot\rho{\bf u} = {\dot \rho}_{hs}
\end{equation}
\begin{equation}
\rho \left( { \partial {\bf u} \over \partial t}
+ ({\bf u \cdot \nabla}){\bf u}\right) =
- {\bf \nabla}(P + P_c) 
+ {\bf \nabla\cdot\Pi }
- \rho {\bf g} + \rho {\bf a}_{hs}
\end{equation}
\begin{equation}
{\partial e \over \partial t}
+ {\bf \nabla \cdot u}e = - P({\bf \nabla\cdot u})
+ {\bf \Pi : \nabla u}
\end{equation}
\begin{equation}
{\partial e_c \over \partial t}
+ {\bf \nabla \cdot u}e_c = - P_c({\bf \nabla\cdot u})
+ {\dot S}_{hs}
\end{equation}
\vskip.1in
\begin{equation}
{\partial {\bf B} \over \partial t} = 
{\bf \nabla \times (u \times B)}~~~
{\bf \nabla \cdot B} = 0.
\end{equation}
Equation (4) for the integrated 
CR energy density does not include 
CR diffusion nor do we include loss terms due to synchrotron or 
inverse Compton emission, assuming that the radiating CRs are a small
fraction of the total CR energy density.
Because of their negligible rest mass, a mass conservation equation for the 
relativistic CR particles is unnecessary.
The viscous stress tensor $\Pi$ appearing in both the momentum 
and internal energy equations is proportional to 
the (assumed spatially uniform) viscosity $\mu$ 
and these terms are provided 
in cylindrical coordinates in the Appendix.
We assume a classical form for the viscous terms, 
but, as discussed below, it must be emphasized that we do not 
understand the physical nature of 
(turbulent or particle) transport processes 
in relativistic, weakly magnetic plasmas 
(Schekochihin et al. 2010). 
Radiative cooling of the thermal gas is not included 
because of the short age of the Cygnus A event, 
assumed to be 10 Myrs.


Our 2D Eulerian code employs advection procedures 
similar to those described by Stone \& Norman (1992) for the ZEUS code, 
but with CR and other features added to study FRII sources.
This code has been used extensively for other similar problems
and correctly duplicates many relevant calculations
such as for example the CR-gas shock structure of 
Jones \& Kang (1990). 
We adopt a modest grid resolution with 
150 uniform zones $\Delta r = \Delta z = 0.5$ kpc out to 75 kpc, 
which completely encloses the 60 kpc size of the Cygnus A cocoon, 
and 50 additional geometrically increasing 
zones in both directions extending out to about 1 Mpc 
into the surrounding cluster gas. 
Our $\sim1$ kpc grid resolution is comparable with the observational 
resolution of 
{\it Chandra} in X-rays and the VLA in radio.

Cygnus A is centered in a large galaxy cluster containing 
a hot gas atmosphere that has been observed in X-rays 
with temperature $kT = 4.60$ keV at 32 kpc
(Smith et al. 2002; Wilson et al. 2006). 
As explained in MG10, to extend these observations of 
the Cygnus A cluster to larger cluster scales, we 
use gas density and temperature profiles 
of the similar cluster Abell 478 (virial mass
$1.25 \times 10^{15}$ $M_{\odot}$ and NFW concentration
$7.61$) from Vikhlinin et al. (2006) 
and renormalize them to agree with the Smith-Wilson observations 
near the cluster center.
The stellar and massive central black hole 
in M87 have been used to represent the 
central galaxy in Cygnus A (see MG10 for further details).

As discussed above and in MG10, 
our computation begins with the moving hotspot, not the jet.
All CRs and magnetic flux that enters the radio lobe 
originates in the hotspot. 
The hotspot is an energy source in two respects, 
as a region where gas and CRs are compressed by 
the jet in the reverse 
shock and as the post-shock region where the jet energy 
is largely isotropized. 

The kpc-sized hotspot observed in Cygnus A is approximated with 
a cylindrical region of radius 1 kpc and height 0.5 kpc  
(two computational zones) 
elongated in the transverse $r$-direction 
as observed at radio frequencies (Carilli et al. 1999).
The velocity of the hotspot $v_{hs}$ can be estimated from the 
age and dimensions of Cygnus A. 
The age $t_a \sim 10^7$ yrs is determined from
the synchrotron lifetimes of radiating CR electrons
(Machalski et al. 2007).
The (projected) distance of the hotspot 
from the center of its host galaxy is 60 kpc.
Assuming uniform motion, the hotspot velocity is 
$v_{hs} = 60~{\rm kpc}/10^7~{\rm yr} = 5870$ km s$^{-1}$,
disregarding a small projection correction onto 
the plane of the sky.
The acceleration and compression of gas in the hotspot 
by the reverse shock occurs in the $z$-direction along the 
jet axis.
The location of the hotspot $z_{hs}(t) = v_{hs}t$ 
at any time determines the two hotspot zones 
$z_{i-{1 \over 2}} < z_{hs}(t) < z_{i+{1 \over 2}}$ 
that receive an acceleration, 
\begin{equation}
a_{hs;i,j} = { \rho_{i,j} v_{hs}(t)^2 A_{i,j}
\over \rho_{i,j} A_{i,j} \Delta z }
= {v_{hs}(t)^2 \over \Delta z}, 
\end{equation}
where $A_{i,j} = \pi(r_{j+1}^2 - r_{j}^2)$
is the area of the two $j = 1,2$ hotspot zones 
and $\Delta z = z_{i-1} - z_i$. 
During each time step $\Delta t = t^{n+1} - t^n$ the 
hotspot velocity at time $t^{n+1}$ is 
\begin{equation}
{u_{i,j}}^{n+1} =
\min[{u_{i,j}}^n + v_{hs}^2 \Delta t/ \Delta z , v_{hs}].
\end{equation}
and $\Delta t$ is chosen so that the gas velocity in the hotspot 
slowly approaches $v_{hs} = dz_{hs}/dt$ over many time steps.

A second, much more important 
hotspot energy source is the CRs introduced 
or created by the jet in the reverse shock at the inner boundary 
of the hotspot. 
In MG10 we found that an average CR hotspot power of 
$L_{cr} = 10^{46}$ erg s$^{-1}$ is sufficient to inflate the 
radio cavity in Cygnus A to approximately 
its currently observed volume $V_{lobe}$. 
In our computation here the hotspot volume $V_{hs}$ remains constant 
as it moves through a uniform grid.  
If $L_{cr}$ is also assumed to be constant, 
the source term in the CR energy density equation is 
\begin{equation}
{\dot S}_{hs} = {de_{c,hs} \over dt} = {L_{cr} \over V_{hs}}
~~~{\rm erg~cm}^{-3}{\rm ~s}^{-1}.
\end{equation}
The addition of CRs to the hotspot contributes a total energy 
$E_{cr} = L_{cr}t_a = 3 \times 10^{60}$ ergs that is 
at least 50 times greater than the work done in accelerating 
the hotspot material to $v_{hs}$.

As indicated with ${\dot \rho}_{hs}$ in equation (1) above,
we also inject a small mass of non-relativistic 
gas into the hotspot at a rate 
1 $M_{\odot}$ yr$^{-1}$, 
that is assumed to arrive with the jet.
This is only about 6 percent of the total rate, 
$\sim17$ $M_{\odot}$ yr$^{-1}$, that 
gas flows into the X-ray cavity 
in our Cygnus A flow calculations (MG10).
Hotspot gas serves to transport the frozen-in magnetic 
field from the hotspot to the radio lobe. 
The thermal energy density of gas inside the hotspot is 
several hundred times smaller than that of the CRs, 
$e_{hs} << e_{cr,hs}$.

In Model 1 of MG10 we also described an evolution of the 
Cygnus A cavity-cocoon with constant 
$v_{hs}$, $V_{hs}$, and $L_{cr}$. 
However, we repeat this calculation again here with 
a slightly different hotspot boundary condition in which 
cosmic rays are not allowed to flow directly from the hotspot 
upstream across the reverse shock 
inside the incoming (but not actually computed) jet.
This new boundary condition only applies to the inner 
boundary of the two hotspot zones and serves to focus 
the kinetic outflow from the hotspot in 
the forward (jet) direction and perpendicular 
to the jet axis. 
CR acceleration by diffusion across the reverse shock 
is sharply reduced by the perpendicular alignment of the 
magnetic field (${\bf B} \bot \bf{\hat{z}}$) 
(Sironi \& Spitkovsky 2009) which is consistent with 
field orientations in Cygnus A observed by 
Carilli et al. (1999).



\section{Non-viscous Computation}

In our first calculation we describe the computed appearance 
of Cygnus A after evolving for $10^7$ yrs 
without viscosity (${\bf \Pi} = 0$).
The bow shock in the cluster gas, the radio cavity and its 
contents are all completely described by the initial cluster 
atmosphere and sources in the 
hotspot, moving out at constant velocity $v_{hs} = 5870$ km s$^{-1}$. 
After each computational time step $\Delta t$ the CR energy density 
in the hotspot is increased according to:
\begin{displaymath}
{e_{c;i,j}}^{n+1} =
{e_{c,i,j}}^n + {\dot S}_{hs} \Delta t
\end{displaymath}
assuming constant ${\dot S}_{hs} = L_{cr}/V_{hs}$. 
The hotspot ($z$-direction) 
grid index $i$ is determined by the 
instantaneous position of the hotspot 
and the $r$-direction grid index $j$ includes both hotspot zones.

Figure 2 (top) shows the density variation $\rho(z,r)$ in 
this Cygnus A calculation at time 10 Myr.
The dark, low-density central cavity 
is surrounded by a denser layer of shocked cluster gas, 
and a bow-cocoon shock separates this region from 
undisturbed cluster gas beyond. 

Superimposed on the X-ray cavity are white contours 
showing the computed distribution of CR energy density $e_c(z,r)$. 
The CR pressure $P_c(z,r)$ and $e_c(z,r)$ 
are nearly constant inside the 
cavity, reflecting the much larger pressure scale height there.
The concentration of white contours near the 
cavity boundary indicates that $e_c(z,r)$,  
and therefore the radio lobe emission, is sharply defined 
as in the observed image (Fig. 1 center).
The concentration of $e_c$ and radio emission near the hotspot 
(at $z = 60$ kpc) is also evident.
Since CRs are not allowed to cross 
the inner boundary of the hotspot zones, this appears as 
a small, sharp vertical $e_c(z,r)$ transition in Figure 2 (top).
Also of interest is the thin, low-density (dark) sheath of  
thermal cavity gas surrounding the outermost contours 
of $e_c$. 
This narrow ``thermal sheath'' may originate in cluster gas 
just beyond the hotspot (Section 4.2), 
but higher-resolution computations 
would be useful to verify the origin of this feature
which contains most of the thermal gas mass 
inside the cavity.

However, the outer boundary of the radio emitting 
region shown with white contours in Figure 2 (top) has 
large folds associated with vortical flows inside the 
radio cavity. 
Evidently these are produced by the Kelvin-Helmholtz (KH) 
instability expected in the rapidly shearing backflow. 
Such KH features, also appearing in the flows computed in MG10, 
have not been observed in radio frequency 
images.

Figure 2 (center) shows a a superposition of gas velocity 
flow vectors for this solution.
The largest flow speed, $0.48c$, 
which occurs just as hot gas and CRs emerge 
from the hotspot, is mildly relativistic 
$(|{\bf u}|/c)^2 = 0.23$, but
$|{\bf u}|/c$ decreases rapidly in neighboring grid zones.
As a result of the boundary condition on the 
two hotspot zones 
that forbids gas from moving upstream 
into the oncoming jet, 
CRs and gas exit from all other directions. 
Shortly thereafter, the hotspot outflow 
is redirected back along the surface of 
the radio cavity. 
Unlike many, possibly all, previous FRII calculations, 
our backflow lies just along the radio lobe boundary, 
subsequently referred to as the  ``boundary backflow''. 
However, three large vortices at $z = 6$, 24 and 36 kpc 
visible in Figure 2 (center) -- 
presumably due to KH instabilities -- 
cause CRs of different ages to mix together. 
This CR mixing upsets the smooth variation of 
synchrotron ages apparent in FRII observations (Fig. 1 bottom). 
This is another major drawback of this flow evolution.

Figure 2 (bottom) shows 
contours of the time in Myrs since  
the local gas and CRs left the hotspot.
To evaluate this time, we use a new variable formed 
by the product of the gas density and time in Myrs,
$\tau = \rho t$. 
At each time step in the calculation the gas density 
in the hotspot and the current time are used to define 
$\tau$ which then advects away from the hotspot 
just like the gas density,
\begin{equation}
{ \partial \tau \over \partial t}
+ {\bf \nabla}\cdot\tau{\bf u} = 0.
\end{equation}
At any point in the subsequent flow inside the cavity 
the time at which the local gas exited the hotspot 
can be found by division, $t_{exit}(z,r) = \tau/\rho$. 
We use this time to determine the local radio synchrotron age 
$t_{age}(z,r) = 10 - \tau/\rho$ Myr. 
The contours in Figure 2 (bottom) describe the locus of the 
mean emission-weighted age (in Myrs) of CR electrons at each 
position in the radio lobe,
\begin{displaymath}
\langle t_{age} \rangle = 
{\int e_c B^2 (10 - \tau/\rho)ds \over
\int e_c B^2 ds}
\end{displaymath}
where $e_cB^2$ is an approximate 
surrogate for the synchrotron emissivity 
and the evolution of the magnetic field is discussed in 
Section 4.1 below.

In Figure 2 (bottom) it is seen that 
vortical KH and other dynamical activity inside 
the radio cavity has mixed 
$\langle t_{age}(z,r) \rangle$ 
in a way that is inconsistent with the 
characteristic distribution 
of radio synchrotron ages in FRII sources  
which varies smoothly and monotonically 
across the radio lobes as in Figure 1 (bottom). 
The computed apparent age distribution in Figure 2 (bottom) 
is a mess. 
The vortical flow inside the lobe has distorted the 
contours of 9 Myr old CRs and sundered the 8 Myr old 
CRs into two separate regions -- we are unaware 
of any FRII observation that shows such strong  
and prominent CR age oscillations along the radio lobe.

Figure 3 shows profiles perpendicular to the jet 
direction ($z$-axis) 
of the flow velocities and the time when local 
gas and CRs left the hotspot. 
While the negative boundary flow at $z = 55$ kpc 
(closest to the hotspot)
has a well-defined time, profiles further back along 
the radio lobe become strange: at $z = 35$ kpc the flow 
at $r \lta 5$ kpc is strongly positive, 
and at $z = 25$ kpc the backflow near the boundary 
has reversed, becoming positive 
due to a local vortex.
This again is evidence of the failure of this  
flow to conform to the observed distribution 
of radio synchrotron ages. 

As a result of KH instabilities in 
the evolution computed in Figures 2 and 3,  
this model fails to match the Cygnus A observations 
in two important ways: (1) the outer radio lobe 
contours are irregular due to vortical folding and 
(2) the CR age distribution is rather drastically 
upset and rearranged.
The chaotic radio cavity flow in 
Figure 3 apparently also occurs in many other 
FRII computations such as the recent 3D models 
of Hodges-Kluck \& Reynolds (2011) and Huarte-Espinosa, Krause \&
Alexander (2011).  
Neither the cavity kinematics nor the 
distribution of radio ages of 
synchrotron electrons are discussed in detail by these authors, 
but their images of highly irregular FRII cavity
boundaries and internal structures necessarily 
require disordered and disruptive velocity fields inside the
cavities. 
By comparison, the radio images of FRII sources are much 
smoother and regular 
(e.g. Kharb et al. 2008). 
Nevertheless, the width of the observed Cygnus A radio lobe 
in Figure 1 (bottom) does not vary in a perfectly 
monotonic fashion along the lobe. 
But these variations in the Cygnus A lobe width 
are not accompanied by strong vortical
mixing of CR electron ages.
A locally smaller lobe width could
arise for example from a past asymmetric hotspot excursion
or a decrease in hotspot CR production during a time in the past.
In any case the model we compute in Figures 2 and 3, 
and those of other authors, are unsatisfactory. 
We conclude that KH instabilities must be damped.

\subsection{Stabilizing KH: magnetic field or viscosity?}

The two most plausible ways to remove the KH 
instability in the radio cavity 
are magnetic tension or viscous damping.
Magnetic KH stabilization is possible if 
the field has significant non-toroidal components and if 
the relative backflow velocity
along the bubble surface is less than the root mean squared average of the
Alfv\'en speeds $B/(4 \pi \rho)^{1/2}$ on both sides 
of an interface (Chandrasekhar 1961).
This second 
condition cannot be satisfied with the magnetic fields observed 
and our computed density. 
Consider for example the flow near the cavity boundary 
in the previous solution. 
At $z = 55$ kpc the lobe boundary backflow velocity decreases 
from a maximum of $|{\bf u}| = 9.0 \times 10^4$ km s$^{-1}$ at $r = 4$ 
kpc to
$|{\bf u}| = 1 \times 10^4$ km s$^{-1}$ at $r = 5.5$.
Across this same region the 
gas density increases from $2 \times 10^{-5}$ to $10^{-4}$ cm$^{-3}$
and the Alfv\'en speed decreases from 
$v_A = 1 \times 10^4$ km s$^{-1}$ to
$v_A = 0.3 \times 10^4$ km s$^{-1}$, 
when evaluated with the maximum observed lobe field 
$B \approx 20\mu$G (Yaji et
al. 2010) and our computed gas density.
Since $v_A < |{\bf u}|$, the observed field is too small to 
stabilize the computed boundary backflow. 
An enormous lobe field, $\sim 200\mu$G, 
comparable with that observed in the 
hotspot, would be required to stabilize the KH instability at this $z$.
(Such a large field would nearly be in equipartition with our 
computed CR energy density $e_c$ inside the radio lobe.)
At $z = 45$ and 35 kpc the computed boundary backflow 
velocities $|{\bf u}| \approx 5 \times 10^4$ km s$^{-1}$ 
are still larger than the Alfv\'en speeds 
$v_A \approx 1.0 \times 10^4$ km s$^{-1}$ 
computed with $20\mu$G, 
as before, $\sim100\mu$G lobe fields would be required for stabilization. 

In addition, magnetic stabilization would require (non-toroidal) fields 
along the lobe boundary roughly in the $z$-direction.
Such large fields would almost certainly have to be inside the 
cavity since fields observed in cluster gas tens of kpc from the center 
are typically very much smaller, only a few $\mu$G
(e.g. Feretti et al. 2009; Bonafede et al. 2010). 
Dursi \& Pfrommer (2008) and Pfrommer \& Dursi (2010)
describe the concentration of tangential fields due to ``draping'' of 
field lines around the upwind-facing surface of 
a moving object.
However, strong field draping does not occur along the backflowing 
boundary of the Cygnus A radio lobe 
since this is a contact discontinuity between two 
fluids, the cavity and shocked cluster gas, 
that are moving at the same velocity 
perpendicular to the interface.

If the gas density inside the radio cavity was 
about 25-100 times lower than we calculate, 
this could increase the Alfv\'en speed to be comparable with 
the backflow velocity. 
The low gas density, $n_e \approx 10^{-4.7}$ cm$^{-3}$, 
that fills most of the boundary backflow
in the non-viscous flow  
arises from the small mass source, $1$ $M_{\odot}$ yr$^{-1}$, 
we introduce into the hotspot.
More detailed future calculations will be necessary to 
determine if extremely low cavity plasma densities 
can increase the Alfv\'en velocity sufficiently 
for magnetic KH stabilization. 

Alternatively, perhaps the magnetic field in 
the radio cavity, $15 - 20\mu$G,
has been underestimated because non-thermal 
emission originates in the boundary backflow 
of volume $V_{bf}$ , not throughout the 
cavity volume $V_{lobe} > V_{bf}$ as commonly assumed
(Yaji et al. 2010).
According to Yaji et al. 
about half of the nonthermal 
X-ray emission comes from inverse Compton (IC) upscattered 
CMB photons and half from synchrotron self-Compton (SSC). 

Assuming for simplicity monoenergetic representative CR electrons, 
the power emitted from a single CR electron 
with mean energy $\gamma m_e c^2$ is 
\begin{equation}
{\dot \varepsilon} = (4/3)(\sigma_T c \gamma^2 \beta^2)u_{ph} \equiv
  a\gamma^2u_{ph} 
\end{equation}
where $\beta = v/c = 1$, $a$ is a constant 
and $u_{ph}$ is the energy density of the relevant 
field of diffuse photons
with space density $N_{ph}$.
The total radio synchrotron luminosity from volume $V$ is 
\begin{equation}
L_{sy} = VN_{sy} \gamma_{sy}^2 a u_B = 
V{\cal N}_{sy} a  {B^2 \over 8\pi}
\end{equation}
where $u_B = B^2/8\pi$,
\begin{displaymath} 
{\cal N}_{ph} \equiv N_{ph} {\gamma_{ph}}^2,
\end{displaymath}
and 
\begin{equation}
N_{sy}  = {e_c\over \gamma_{sy} m_ec^2} = 
{L_{sy} \over Va\gamma_{sy}^2}~{8 \pi \over B^2}
\end{equation}
is the number density of
CR electrons with radio emitting energy $\gamma_{sy}$. 
The IC X-ray luminosity produced by interactions with CMB 
microwave photons 
and CR electrons of energy $\gamma_{ic}m_ec^2$ is  
\begin{equation}
L_{ic} = V N_{sy} \gamma_{ic}^2 a u_{cmb}
= {{\cal N}_{ic} \over {\cal N}_{sy}}L_{sy} 
{u_{cmb} \over u_B}.
\end{equation}
This last equation indicates that the magnetic field 
found from X-ray IC observations
\begin{displaymath}
B \approx \left(8 \pi {{\cal N}_{ic}\over {\cal N}_{sy}} 
{L_{sy} \over L_{ic}} u_{cmb}   \right)^{1/2}
\end{displaymath}
is independent of the volume.
For SSC X-rays the photon energy density is 
$u_{sy} = (L_{sy}/V)(r/c)$ where $r$ is the appropriate 
mean dimension across $V$ along which photons are upscattered.
The SSC luminosity is 
\begin{equation}
L_{ssc} = V N_{ssc}{\gamma_{ssc}}^2 a u_{sy} =
{{\cal N}_{ssc}\over {\cal N}_{sy}} L_{sy} {u_{sy} \over u_B}
= {{\cal N}_{ssc}\over {\cal N}_{sy}} {L_{sy}}^2 {(r/c) \over Vu_B}.
\end{equation}
This last equation indicates that the field determined from 
X-ray SSC emission varies as $B \propto (r/V)^{1/2} \sim 1/r_{eff}$
where $r_{eff}$ is the mean distance characterizing 
the SSC emitting volume.
If Yaji et al. had considered SSC emission from the smaller 
boundary flow volume, 
$r_{eff}$ would be reduced by less than a 
factor of about two.
Evidently, the increase in the estimated $B$ 
by using $V_{bf}$ rather than $V_{lobe}$ does not 
provide sufficient magnetic tension to stabilize 
KH instabilities in the computed backflow.
Apart from CR energy dependent factors, 
the non-thermal luminosities have the following dependencies:
\begin{displaymath}
L_{sy} \propto e_c u_B V
\end{displaymath}
\begin{equation}
L_{ic} \propto e_c u_{cmb} V
\end{equation}
\begin{displaymath}
L_{ssc} \propto e_c u_{sy} V = {e_c}^2 u_B V(r/c)
\end{displaymath}
where $u_B = B^2/8\pi$ and 
$u_{sy} = L_{sy}(r/c)/V$ 
is the energy density of synchrotron radiation.

Aside from these concerns,
an additional preference for damping KH instabilities 
with viscosity rather than magnetic fields 
is motivated by our successful application of 
viscous damping to the Fermi bubbles in the 
Milky Way (Guo et al. 2012), where the magnetic fields that 
could provide similar KH damping seem prohibitively large.

Nevertheless, magnetic fields are directly observed 
in FRII sources, 
while there is less compelling observational evidence for viscosity, 
although it has been invoked to explain the smooth boundaries 
of X-ray cavities in cluster gas (Reynolds et al. 2005).
The classical kinematic particle viscosity is 
$\mu \propto (1/3){\bar v}m_p\lambda$ 
where ${\bar v}$ is the mean random velocity of protons 
of mass $m_p$ with mean free path 
$\lambda$. 
The non-magnetic (Spitzer) plasma viscosity is 
$\mu = 5500 (T/10^8 {\rm K})^{5/2}$ gm cm$^{-1}$ s$^{-1}$, 
but the corresponding mean free path 
$\lambda \approx 20 (T/10^8 {\rm K})^2 (n_e/10^{-3} {\rm cm}^{-3})$
kpc is largely irrelevant in a magnetized plasma 
because particle excursions are constrained by 
the much smaller Larmor gyroradius, 
$r_g = 1.5 \times 10^8 (T/10^8 {\rm K})^{1/2}(B/20\mu{\rm G})^{-1}$ cm
for a thermal proton.
The viscosity that we require to transport momentum 
in the boundary backflow 
may therefore result from localized turbulent activity rather than 
particle trajectories.
But turbulence must 
not mix CR electrons of different ages on large scales -- 
this may be a difficult constraint to satisfy.
We also recognize that shear viscosity in the radio lobes 
may be incompatible with our assumption of purely 
toroidal fields aligned perpendicular 
to the shear gradient.
However, we note that there is evidence 
in the Virgo cluster that energetic particles can diffuse 
across the observed field direction, as required to 
produce the bright rims on large cluster gas radio lobes
(Mathews \& Guo 2011). 
Cross-field diffusion and momentum transport 
may be possible due to small-scale field irregularities 
and MHD turbulence that are difficult to observe.
In any case, for the exploratory calculations here 
we invoke viscosity to stabilize the KH instability.
Observations require that 
the KH kinematic features in Figure 2 must be removed and it 
may not matter whether this is accomplished with 
magnetic fields or viscosity. 


\section{Viscous Computation}

We now repeat the previous computation 
but include a non-zero shear viscosity.
For simplicity the solution we describe has a uniform 
viscosity $\mu = 30$ gm cm$^{-1}$ s$^{-1}$ throughout 
the computational grid. 
This value is sufficiently small not to have any appreciable
effect on the flow outside the radio cavity where shear velocities 
are much smaller. 
We also considered a somewhat larger value, 
$\mu = 100$ gm cm$^{-1}$ s$^{-1}$,
but found that the boundary backflow was broader 
than observations allow, as discussed below. 

The density variation of this flow, 
illustrated in Figure 4 (top), 
shows that the KH features in Figure 2 are largely absent.
Only one rather weak vortical KH feature 
appears in Figure 4 within about 5 kpc from the center, but the 
CR energy density $e_c$ contours are now very regular 
throughout the cavity interior.
The central panel shows that the boundary backflow 
now flows smoothly without interruption nearly to 
the $z = 0$ plane. 
The much slower gas flow in the central region 
of the cavity is in the opposite sense, 
along the direction of the original jet.

The bottom panel of Figure 4 shows contours of the
local mean line of sight CR ages 
$\langle t_{age}(z,r) \rangle$ at 9, 7, 5 and 4 Myrs 
since leaving the hotspot.
Unlike the non-viscous flow in Figure 2 (bottom), 
the CR ages are now distributed monotonically 
along the radio cavity just as observed 
in Figure 1 (bottom). 
Also in general more youthful CRs are found 
further back along the radio cavity than in the non-viscous 
solution.
Furthermore, the apparent CR age distribution in 
Figure 4 (bottom) often increases along perpendicular  
directions away from the jet, 
as clearly seen in Figure 1 (bottom).
The maximum velocity in the cavity, $u = 0.48c$,
appears in the computational zone 
just adjacent to the hotspot but
declines very rapidly to $\lta 0.1c$ beyond about
2 kpc from the hotspot.

Figure 5 shows more quantitatively 
the correspondence of transverse 
profiles of the velocities ($u_z,u_r$)
and hotspot exit times $\tau/\rho$ at four distances 
along the cavity: 
$z = 55$, 45, 35, and 25 kpc. 
By comparison with Figure 3 it is seen that 
the boundary backflow velocities decrease somewhat faster 
when viscosity is included. 
More importantly, 
no KH vortices interfere with the boundary backflow 
or the monotonic spatial variation of CR ages. 
Viscous momentum diffusion also 
causes the width of the boundary backflow to 
increase downstream from the hotspot.
Even at $z = 10$ kpc the boundary backflow 
still has a large negative velocity,
$u \approx -2300$ km s$^{-1}$,
extending over $4 < r < 12$ kpc.

\subsection{Magnetic Field Evolution}

Detailed radio polarization observations of Cygnus A (Carilli et
al. 1999) indicate that the prevailing sense of magnetic field
orientation in the Cygnus A hotspots is parallel to the shock plane.
This is consistent with a large toroidal field component but
does not exclude hotspot fields with radial $r$-components ($B_r \ne
0$).  Nevertheless, for simplicity in this exploratory calculation we
assume that the field is purely toroidal ($B_r = B_z = 0$).
Both equations (5) for the passive frozen-in field evolution 
are satisfied in the toroidal case by solving
\begin{equation}
{\partial B \over \partial t} =
-{\partial \over \partial z}(B u_z)
-{\partial \over \partial r}(B u_r)
\end{equation}
since toroidal fields $B \equiv B_{\phi}$ are 
inherently divergence-free.
At each time step the 
toroidal magnetic field in the hotspot
is reset to $B_{hs}$ and the continuous wind outflow from the
hotspot transports the field throughout the radio cavity, 
preserving its toroidal morphology.
All magnetic fields in the radio lobe 
are assumed to originate in the hotspot. 
However, in the Cygnus A radio lobe downstream 
from the bright hotspots 
the field orientation indicated by 
radio polarization observations 
is poloidal, often parallel to the lobe boundary 
(Perley \& Carilli 1996), so some original poloidal 
field component must also be present in the hotspot
\footnotemark[2]. 
A poloidal field component in the hotspot would be amplified
by transverse shear in the boundary backflow.

\footnotetext[2]{
Even if the hotspot magnetic field were exactly 
toroidal relative to the jet direction, 
the instantaneous jet and hotspot positions are not symmetric 
with the radio lobe symmetry axis (Fig. 1 center), 
so production of poloidal fields is unavoidable.
}

For our limited purposes in this paper 
an idealized purely toroidal field serves as 
a convenient approximate tracer of field evolution 
as the field advects downstream back into the cavity. 
Strong compression by the jet-hotspot shock would realign 
a random magnetic field until only field components 
parallel to the shock plane survive 
(Laing 1980).
The fractional radio polarization observed 
by Carilli et al. (1999) in the 
bright central cores of the Cygnus A hotspots is $\sim20-30$\%, 
indicating that the field alignment,
although high, is not that of 
a purely toroidal field 
in which the degree of polarization would 
peak at $\sim70$\% near the hotspot center. 
The field in the hotspot is likely to 
be a compressed version of the field arriving 
in the jet perhaps with additional non-linear 
amplification driven by post-shock instabilities and 
magnetic turbulence. 
In addition, 
it is unclear how much of the total hotspot emission 
and apparent field orientation 
comes from small 10pc-sized hotspot inhomogeneities 
visible with VLBI observations 
(Tingay et al. 2008). 

The toroidal field assumption allows us 
for the first time to 
relate the field strength observed
in the Cygnus A hotspots,
$B_{hs} \approx 170-270\mu$G (Stawarz et al. 2007) 
with much smaller fields observed in the radio lobes,
$B_{lobe} \approx 15-20\mu$G (Yaji et al. 2010).
A comparison of Figures 3 and 5 illustrates how well 
viscous damping organizes the CR flow inside the radio cavity.
While the lobe field is 
reduced by adiabatic spatial expansion into the 
large cavity volume, the lobe field is 
amplified by compressive deceleration along the boundary backflow
(see below). 
This latter point makes it difficult to reconcile 
the observed hotspot and lobe field strengths.
For example, 
assuming a hotspot field of $B_{hs} = 200\mu$G, 
the computed fields at time 10 Myrs 
in the non-viscous boundary backflow 
in the range $ 25 \lta z \lta 45$ kpc 
would have maximum values $590 \gta B \gta 400\mu$G, 
more than 20 times the observed values.
These fields have maximum energy densities 
$1.4 \times 10^{-8} \gta B^2/8\pi \gta 6.4 \times 10^{-9}$ 
erg cm$^{-3}$ 
that exceed the 
CR energy density in the lobe 
$e_c \approx 2 \times 10^{-9}$ erg cm$^{-3}$. 
Such large super-equipartition fields, which would rapidly 
evolve toward a force-free morphology, are most unlikely.

The relatively small volume-averaged 
radio cavity field, $B_{lobe} \approx 15-20\mu$G,
is found by interpreting 
the X-ray emission from the 
(outer half of the) lobes as a combination of 
IC-CMB and SSC in approximately equal proportions 
(Yaji et al. 2010).
This is a more straightforward field estimate than that 
in the hotspot where post-shock conditions are uncertain 
and nonlinear inhomogeneities within hotspots are often observed 
(e.g. Carilli et al. 1999; 
Tingay et al. 2008;
Wright \& Birkinshaw 2004;
Hardcastle, Croston \& Kraft 2007; 
Perlman et al. 2010; Erlund et al. 2010; Orienti et al. 2011). 
The synchrotron emission in the hotspots
may be concentrated in small inhomogeneous regions of high
field energy $B^2/8\pi$ and/or high CR electron density,
that skew the estimated mean hotspot field to higher values.
Moreover, Weibel-type instabilities or other
non-linear post-shock transient fields
(e.g. Tatischeff 2008) may generate strong
localized post-shock fields that later decay. 
Consequently, for the purposes of our approximate field 
calculation we adopt a smaller hotspot field 
$B_{hs} = 60\mu$G, chosen to avoid $u_B = B^2/8\pi > e_c$
in the radio lobes. 
The passively advecting post-hotspot field we compute 
with equation (16) can be linearly rescaled to any
other suitable hotspot field $B_{hs}$ that is consistent with our
neglect of the Lorentz force. 

Figure 6 shows a superposition of transverse profiles 
of the CR energy density $e_c$ and the toroidal magnetic field 
energy density $u_B = B^2/8\pi$ 
at $z = 55$, 45, 35, 25 (heavy lines) 
and 5 kpc (thin lines). 
The upper and lower panels 
show results without and with viscosity respectively.
The CR energy density is approximately uniform 
throughout the lobe as expected. 
However, the magnetic field energy density $u_B$ 
is seen to peak strongly near the 
outer edge of the boundary backflow. 
In the viscous solution the maximum field strength at 
$z = 55$, 45, 35 and 25 is 
$B = 50$, 110, 160, and 210$\mu$G respectively. 
At $z = 25$ kpc $u_B \approx e_c$ and Lorentz forces would 
become important if our adopted hotspot field normalization 
($B_{hs} = 60\mu$G) were correct. 

A more serious concern with Figure 6 is that even 
our adopted hotspot field $B_{hs} = 60\mu$G, 
which we have lowered in an ad hoc manner, 
results in radio lobe fields that are 
too large for the expected synchrotron lifetimes. 
If the mean field in the outer part of the radio 
lobes ($z \gta 35$ kpc) is $B \approx 100\mu$G 
($B^2/8\pi \approx 4 \times 10^{-10}$ erg s$^{-1}$),
the corresponding synchrotron lifetime,
$t_{sy} = 1.41 \times 10^9 
{\nu_{\rm GHz}}^{-1/2}{B_{\mu{\rm G}}}^{-3/2}$ yrs,
for radiation 
at 5 GHz is only $\sim 10^6$ yrs. 
By contrast, electrons emitting at this 
same frequency in the $20\mu$G radio lobe magnetic field measured 
by Yaji et al. (2010) have a more reasonable 
lifetime, $t_{sy} \approx 10^7$ yrs. 
Overall, with a hotspot field $B_{hs} \sim 200\mu$G 
the computed lobe field is about 20 times 
larger than observed.  

Many possible explanations 
for the discrepancy between the hotspot 
and lobe fields can be imagined:
(i) difficulty in estimating the mean hotspot field due 
to inhomogeneities and local field gradients, 
(ii) the field estimate of 
Yaji et al. can be increased if the effective 
volume that radiates SSC X-rays was overestimated, 
(iii) poloidal field components in the hotspot may  
produce smaller fields in the lobe backflow. 
However, if the hotspot field 
was random, the toroidal hotspot component would be reduced 
from our value by only $3^{-1/2}$ and would evolve 
independently of the poloidal field components, 
resembling Figure 6, 
(iv) if viscous turbulence is the agency that stabilizes KH 
instabilities, the orientation of the magnetic field 
inside the lobes would be more random, possibly 
resulting in field loss by reconnection, 
(v) perhaps CR electrons are accelerated by turbulence 
in the lobe environment so their synchrotron lifetime 
$t_{sy}$ can be less than the flow time from the hotspot,
(vi) the hotspot field may increase with time, etc.
In any case, the substantive conclusions we 
discuss below are unaffected by the precise magnitude 
of the radio lobe field, assuming Lorentz terms can 
be ignored.

Of particular interest in Figure 6 
is the general increase in magnetic energy density 
with increasing distance from the hotspot due to compression 
in the decelerating boundary backflow seen in 
Figure 5  -- the field follows the gas flow,
not its density.
According to equation (16), 
as an initially uniform backflow ($\partial B/\partial z = 0$)
in the $z$-direction 
decelerates ($\partial u_z /\partial z < 0$),
the field grows exponentially 
$B \propto \exp(|\partial u_z/\partial z|t)$, 
assuming that the hotspot field $B_{hs}$ is roughly 
constant with time. 
Such spatial field variations 
have not been considered in estimates of mean 
fields in FRII sources, although 
gradients in $u_B$ might be detectable 
by comparing radio synchrotron and X-ray (IC-CMB and/or SSC) 
emission from different parts of the radio lobe.
If the field increases along the boundary backflow 
as we propose, this 
would also influence estimates of the radio-synchrotron age 
as a function of distance from the hotspot. 
Finally, in Figure 6 the field is seen to 
essentially disappear near the jet axis.
The similarity of the two panels in Figure 6 indicates that 
viscosity has little to do with the post-hotspot evolution of 
the field.

Aside from difficulties normalizing 
the magnetic field in the lobe with that in the hotspot, 
the field variations in the lobe we 
compute have a morphology entirely consistent with 
radio observations 
of Cygnus A and other similar FRII sources.
Following equation (12), we expect the 
radio lobe synchrotron emissivity $L_{sy}/V$ 
to vary as the product 
of CR and magnetic energy densities, $e_cu_B$.  
Likewise, the radio surface brightness distribution 
in the lobe is proportional to the integral of the emissivity  
$e_cu_B$ over the line of sight, $\int e_cu_Bds$.
In Figure 7 (top) we plot surface brightness 
(solid lines) and corresponding 
emissivity profiles (dashed lines) transverse to the jet direction 
at two distances $z = 35$ and 45 kpc. 
Although the surface brightness is fairly uniform with $r$,
it is seen that the emissivity is strongly peaked toward 
the boundary backflow and is not uniform throughout the 
radio lobe as commonly assumed.
This limb-brightened emission pattern closely matches observations 
of Cygnus A shown at the bottom of Figure 7 (Carvalho et al. 2005)
and many other FRII sources (Daly et al. 2010),
clearly indicating that many FRII lobes are radio-hollow. 
This is an excellent overall confirmation of the 
boundary backflow we calculate for Cygnus A.
[A similar emitting boundary backflow can explain the 
nearly uniform gamma ray surface brightness across  
the Fermi bubbles in the Milky Way (Guo et al. 2012) 
as observed by Su, Slatyer, \& Finkbeiner (2010).]

Regarding the discrepancy between field strengths computed 
in the Cygnus A hotspots and lobes, suppose shear in and near
the backflow is sufficient to amplify non-toroidal
fields to be in equipartition with the CR energy density,
$B_{eq} = (8 \pi e_c)^{1/2} \sim 250\mu$G or larger. 
This could be a problem with the lower panel in Figure 7 
since the Lorentz force of poloidal fields comparable to $B_{eq}$
would further broaden the width of the computed backflow, 
conflicting with the backflow widths
observed by Carvalho et al. (2005) in addition to 
the small lobe fields estimated by Yaji et al. (2010).

\subsection{Radio--X-ray Offset in Hotspot}

High resolution VLA and Chandra observations of FRII hotspots indicate
that radio hotspots are often offset several kpc further 
along the jet direction from the X-ray hotspots 
(Hardcastle, Croston \& Kraft 2007;
Kataoka et al. 2008; Perlman et al. 2010;
Erlund et al. 2007, 2010).  
Similar radio-X-ray offsets have been observed in 
knots (internal shocks) in radio jets as in 
Cen A (Hardcastle et al. 2003) and elsewhere.

But radio-X-ray offsets are not observed in the four hotspots in
Cygnus A (Wilson et al. 2000). 
Although the currently observed jet
has moved away, no longer impacting any of the 
hotspots (Fig. 1 center) -- it must be recognized 
that an abruptly redirected jet leaves behind a jet fragment
that continues to impact the old hotspot. 
Hotspots without jet excitation should decay rapidly 
in a sound-crossing time, only $\sim3\times 10^4$ yrs.
Multiple hotspots in FRII sources are common, 
suggesting that the jet direction changes abruptly through
small angles. 
Apparently the jet direction in FRII sources does not move slowly, 
as expected for example by black hole precession,  
since multiple hotspots are often widely separated 
and each hotspot is highly concentrated, lacking comet-like extensions.
Evidently, the re-directed,
spotless current jet (Fig. 1 center) has not yet reached dense cluster gas
at the inner boundary of the radio cavity. 

However, the moving hotspots in our computed FRII flows 
are by design continuously activated by the (virtual) jet 
and they are capable of producing radio-X-ray offsets.
To illustrate this, in Figure 8 we show the projected appearance 
of the hotspot region at time 10 Myrs approximately as 
it would currently appear at X-ray IC-CMB and radio frequencies. 
Figure 9 shows the variation of flow parameters along the jet 
direction in the hotspot region.
From equations 10-15 and Figure 9 the X-ray IC-CMB emissivity
($L_{ic}/V \propto e_c u_{cmb}$) 
is expected to peak at the hotspot with
the local energy density of CR electrons $e_c$.
The light line contours in Figure 8 
show the CR energy density integrated along the 
line of sight, $\int e_cds$.
These contours, peaking at the hotspot 
position ($z_{hs} = 60.5$ kpc), characterize the 
surface brightness distribution of 
IC X-ray emission from upscattered CMB photons 
inside the hotspot wind.
The radio synchrotron surface brightness 
is represented with a similar 
integral over our surrogate for  
the synchrotron emissivity, $\int e_cu_Bds$, 
shown in Figure 8 with heavy line contours peaking 
in an arc-shaped region ahead of the hotspot at 
$z = 61.75$ kpc.
If the Cygnus A hotspots were currently activated by a jet, 
our computed flow would predict a 1-1.5 kpc offset of 
radio from X-ray IC-CMB emission.

The emissivity of X-ray SSC radiation $L_{ssc}/V \propto e_c u _{sy}$
depends on the energy density of synchrotron radiation 
$u_{sy} \approx L_{sy} (r/c)/V$ 
which is non-local 
and should be calculated with an integration over 
the entire radio-emitting region. 
In view of its more complex non-local nature,
we do not plot the SSC X-ray surface brightness in Figure 8.
However, guided by the approximate relations 10-15 and Figure 9,
SSC X-ray emission may not peak at the hotspot 
($z_{hs} = 60.25$ kpc) like IC-CMB, 
but much closer to the radio-synchrotron arc 
($z_{arc} = 61.75$ kpc).
The energy density in synchrotron radiation on $r\sim1$ kpc 
scales inside a volume $V \approx 4\pi r^3/3$
centered at the arc-emitting region in  
the hotspot wind is 
$u_{sy} \approx 3L_{sy}/4\pi c r^2 \approx 1.3\times 10^{-11}$
erg cm$^{-3}$ where 
$L_{sy} \approx \nu F_{\nu}4\pi d^2 = 1.6 \times 10^{43}$ 
erg s$^{-1}$, $d \approx 230$ Mpc and 
$\nu F_{\nu} \approx 2.5 \times 10^{11}$ Jy Hz  
(Stawarz et al. 2007).
If most of the radio synchrotron emission comes from the 
arc offset region, the radius of the arc emission is about 
a third of the hotspot-arc offset distance. 
If so, the synchrotron 
energy density at the hotspot should be lower,
$u_{sy,hs}/u_{sy,arc}\sim1/9$. 

To produce X-ray photons of energy $h\nu_X  = 1$ keV  
from IC-CMB upscattered CMB photons of typical energy 
$h\nu_{cmb} \approx 1.1 \times 10^{-15}$ erg, 
we require CR electrons with energy 
$\gamma_{ic} = [(3/4)h\nu_X/h\nu_{cmb}]^{1/2} = 1060$.
Likewise, photons of frequency 
$\nu_{sy} = 3$ GHz near the radio peak in Cygnus A 
(Figs. 5 and 6 of Stawarz et al. 2007) require CR electrons 
of energy $\gamma_{ssc} = 7700$ to upscatter to 1 keV.
From equations 13 and 14
the ratio of SSC and IC-CMB X-ray emission from the 
arc region is 
\begin{displaymath}
{L_{ssc,arc} \over L_{ic,arc}} = 
{{\cal N}_{ssc,arc}\over{\cal N}_{ic,arc}} \cdot
{u_{sy,arc} \over u_{cmb}} \approx 0.57 \cdot 32 \approx 18
\end{displaymath}
where ${\cal N}_{ssc}/{\cal N}_{r} \approx 0.57$ 
is found from the CR electron energy distribution 
(Fig. 8 of Stawarz et al 2007).
The ratio of IC-CBM X-ray emission from the hotspot region 
to IC-CMB from the arc is 
\begin{displaymath}
{L_{ic,hs} \over L_{ic,arc}} \approx 
{e_{c,hs}\over e_{c,arc}} \approx 1.7
\end{displaymath}
where we consider the same volume $V$ in both 
regions and assume $e_c \propto N_{ic}$. 
The ratio of SSC X-ray emssion between the hotspot and arc
is roughly 
\begin{displaymath}
{L_{ssc,hs} \over L_{ssc,arc}} 
\approx {e_{c,hs} \over e_{c,arc}}
{u_{sy,hs} \over u_{sy,arc}} \approx 
1.7 \left({1 \over 9} \right) \sim 0.19
\end{displaymath}
X-ray emission from SSC is expected to appear throughout
the hotspot-arc region while IC-CMB is more concentrated
in the hotspot,
so we expect the total nonthermal X-ray emission 
to incerase somewhat toward the radio arc,
\begin{displaymath}
{(L_{ssc} + L_{ic})_{hs} \over (L_{ssc} + L_{ic})_{arc}} 
\approx {(2+1) \over (18+1)} {L_{ic,hs}\over L_{ic,arc}}  \sim 0.26.
\end{displaymath}

These approximate considerations suggest 
that the most intense nonthermal X-ray emission is
SSC near the radio-synchrotron arc, 
about 10 times brighter than  
IC-CMB X-rays from the hotspot. 
This may explain why radio-X-ray offsets   
in Cygnus A are small or difficult to observe.  
Other FRII sources with more pronounced radio-X-ray offsets 
may require $L_{ssc}/L_{ic} < 1$. 
Furthermore, the $\sim200\mu$G magnetic field, 
estimated for Cygnus A by Stawarz et al. (2007) 
from a one-zone radio and X-ray SSC emission model, 
probably refers to the enhanced field in the offset region 
as we have assumed. 
In our calculation the field we impose 
in hotspot grid zones is about half that in the 
arc region, but this field reduction of $\sim2$ 
provides only a partial 
explanation of the hotspot-lobe field disparity 
in our dynamical model.

The physical origin of the offset lies in the powerful outflowing 
wind from the hotspot which is the source of  
both CR and magnetic energy for the entire radio cavity.
X-ray IC-CMB emission is expected to peak in the 
hotspot where $e_c$ 
and Compton scattering are greatest at any time.
The CR energy density inside the computed hotspot, 
$e_{c,hs} \approx 2 \times 10^{-8}$ erg cm$^{-3}$,  
is required to inflate the Cygnus A cavity volume 
during the 10 Myr age of the FRII event. 
This hotspot CR energy density is about 
ten times larger than the total CR electron energy density  
required by Stawarz et al. (2007) to explain 
non-thermal hotspot emission; 
this argues against electron pair dominance and 
implies an additional component of relativistic protons.
The CR energy density in our computed hotspot is also 
about ten times larger than the average value of $e_c$ 
in the radio lobe, consistent with a hotspot wind.
The magnetic field introduced inside the high pressure hotspot 
is immediately advected away in the hotspot wind in every direction 
except upstream into the oncoming jet.
As the wind from the hotspot flows in the forward (jet) direction, 
it soon encounters much denser cluster gas behind the bowshock,
The magnetic field advected away from the hotspot in the forward 
flowing wind 
accumulates and increases in strength as the wind decelerates 
toward the dense gas ahead, creating in projection 
an arc-shaped radio hotspot offset ahead of the X-ray hotspot 
as seen in Figure 8.
 


Close-up details of the structure of the hotspot region  
along the $z$-axis are shown 
in low spatial resolution in Figure 9.
The arrow designates the instantaneous hotspot location  
$z_{hs} = 60.25$ kpc at time $t = 10$ Myr.
The hotspot wind extends in the jet direction 
to $z_w = 62$ kpc. 
The pressure in the wind is dominated by CRs, 
$P_c \gg P$, and $P_c$ decreases away from the hotspot.
The wind is subsonic since its kinetic energy density 
is much less than that of the CRs, 
$u_{KE} = \rho {u_z}^2/2 << P_c$.
The magnetic field introduced into the hotspot 
$B_{hs} = 60\mu$G is immediately  
advected away in the wind. 
The field strength in the wind has a small energy density,
$u_{B} = B^2/8\pi \ll P_c$, 
but increases significantly above 
$B_{hs}$ due to deceleration in  
the forward directed wind, i.e. $du_z/dz < 0$. 
At $z_w = 62$ kpc the CR pressure in the wind is balanced 
by the gas pressure in the shocked cluster gas ahead 
so the total pressure $P + P_c$ varies smoothly 
across the wind-cluster gas interface $z_w$. 
Figure 9 shows that 
the gas velocity $u_z$ and density $n_e$ are also continuous at 
$z_w = 62$ kpc with no local change in slope, 
indicating that this interface, close to the 
region of maximum magnetic energy density in the radio peak, 
behaves as a contact discontinuity.

Although our calculation is time-dependent, 
one expects that the narrow region between 
the reverse shock in the hotspot and 
the bowshock ahead does not evolve rapidly 
and is approximately in steady state.
Adopting a simple one-dimensional steady state approximation  
for this intershock flow along the jet axis, we expect that  
the location of the reverse shock in the hotspot 
can be dictated by the location of the bowshock provided 
the Mach number ${\cal M}$ relative to the contact discontinuity velocity
\begin{equation}
{\cal M} = {u_z(z) - u_z(z_w) \over c_s} 
\end{equation}
is everywhere less than unity, where 
\begin{equation}
c_s(z) = \left({[\gamma P(z) 
+ \gamma_c P_c(z)] \over \rho(z)}\right)^{1/2}
\end{equation}
is the local sound speed.
A detailed analysis of our computed flow verifies that 
${\cal M} < 1$ along the jet axis ($r=0$) throughout 
the intershock flow. 
The hotspot wind is also subsonic, which may set 
an interesting lower bound on the density of hot gas 
it contains.
Although it is gratifying that the position of the 
reverse shock and therefore the hotspot can be governed 
by subsonic communication with gas behind the bow shock ahead, 
in our simple computation
the velocity and location of the hotspot are
prescribed in advance.
In a detailed, more realistic FRII computation 
that includes a self-consistent calculation of the jet, 
the reverse shock and 
hotspot location -- and the radio-X-ray offset --  
would be automatically regulated 
by the intershock flow.
Deeper X-ray observations of subsonic hotspot winds
may serve to reveal the physical constituents 
of the hotspot and the jet from which it was created: 
density of non-relativistic gas, 
density of CR electrons or electron pairs,
magnetic energy, etc. 


Offsets of radio emission from X-ray bright hotspots have 
been observed in a number of FRII sources by 
Hardcastle, Croston \& Kraft (2007). 
For example the VLA 8.3 GHz image of  
the brightest western hotspot in 3C 227 is offset 
about 1.3 kpc further along the jet direction 
beyond the {\it Chandra} X-ray hotspot. 
Both hotspots are elongated in directions perpendicular 
to the jet with diameters of $\sim4$ kpc and $\sim7$ kpc
respectively for X-ray and radio spots.
As we show in Figure 8, the computed radio 
hotspot emission extends transversely right 
along the outermost arc-shaped tip of the cavity radio source.

Also encouraging are recent multifrequency observations 
of the incredibly powerful FRII hotspot in 4C74.26
(Erlund et al. 2007; 2010) having a radio--X-ray hotspot 
structure that supports our 
rudimentary offset image in Figure 8.
The X-ray luminous 
4C74.26 hotspot is preceded by an arc of radio emission 
that is offset by 19 kpc (!) in projection.
The large offset distance is 
due in part to the low ambient cluster gas density since the 
4C74.26 hotspot is about $\sim500$ kpc from the cluster center.
The hotspot in 4C74.26 is only detected 
in X-rays while the forward-displaced arc is seen at 
radio, IR, optical and X-ray wavelengths. 
Since no radio emission is observed at the X-ray hotspot, 
Erlund et al. (2010) propose that the  
X-ray emission from the 4C74.26 hotspot is synchrotron in origin,  
probably due to a very energetic CR electron spectrum.
Normally, we expect synchrotron emission to be concentrated 
in the decelerating arc ahead of the hotspot as in Figure 8,
but in 4C74.26 the synchrotron lifetime of X-ray emitting electrons 
with frequency $\nu \approx 2.4\times 10^{17}$ Hz is very short, 
$t_{sy} \approx 
1.4 \times 10^9/{\nu_{GHz}}^{1/2}{B_{\mu{\rm G}}}^{3/2}
\approx 2.9/{B_{\mu{\rm G}}}^{3/2}$ yrs.
This lifetime is very much less than the flow time 
from the hotspot to the radio arc in Figure 9, 
$\sim 5 \times 10^4$ yrs. 
For any reasonable hotspot field 
X-ray emitting synchrotron electrons are necessarily 
confined near their origin in the hotspot core.
X-rays from the leading arc-hotspot 
are CMB photons IC-upscattered by radio synchrotron electrons there. 
In spite of these radiative differences, 
the hotspot morphology observed by 
Erlund et al. closely resembles the offset morphology 
we illustrate in Figure 8.
However, we interpret the arc-shaped radio hotspot as 
a result of an intensified field 
caused by deceleration of a subsonic hotspot wind, 
not necessarily an additional (third) shock 
as proposed by Erlund et al. (2010).

\subsection{Thermal X-ray Jet-Filament and Sheath}

The solid line profile in Figure 10 (top) 
shows a narrow thermal gas density concentration along 
the jet axis.
A smaller density is seen at $r = 4.75$ kpc,  
but at $r \gta 5$ kpc the gas density in the 
radio cavity is about three percent of that in the original 
cluster atmosphere (upper dashed line). 
Density profiles transverse to the jet direction 
are shown in Figure 10 (bottom) 
with central peaks due to the radial filament.

A similar faint radial X-ray feature 
is also observed in Cygnus A 
just along the symmetry axis (Figure 1 top). 
An X-ray spectrum of the 
brighter eastern half of this X-ray filament 
by Steenbrugge, Blundell \& Duffy (2008)
is found to have a non-thermal spectrum 
with photon spectral index of 1.7, 
possibly with  
a small additional thermal contribution. 
On the basis of this observation, which includes 
all X-ray emission from the cavity and cluster 
along the line of sight,
these authors conclude that this diffuse X-ray filament   
is the non-thermal relic of a previous jet. 
However, they did not compare the emission from the 
filament with its immediate cavity environment 
which, according to Yaji et al. (2010), is filled 
with SSC and IC-CMB X-radiation with a nearly 
identical spectral index 1.7. 
Yaji et al. explicitly exclude  
the X-ray filament and hotspots 
from their spectral analysis. 
Consequently,  
the interpretation of this faint linear X-ray feature 
as a dying jet is less compelling because it 
is immersed in non-thermal emission having a similar 
X-ray spectrum. 
Moreover, there is no comparable 
X-ray emission from the current, more youthful 
western radio jet visible in Figure 1.

We propose instead that this X-ray filament 
is a mid-cavity upflow of thermal gas 
similar to those computed in other X-ray cavities 
(e.g. Mathews \& Brighenti 2008).
An outward stream of (low-entropy, 
relatively high-metallicity) 
thermal gas from the cluster core 
is a natural reaction to a transient
symmetric cavitation in cluster gas\footnotemark[3]. 
After a dynamical time, a few $10^8$ yrs, 
this filament will fall back to the cluster center with 
a turnaround point that moves upward along the filament
(Mathews \& Brighenti 2008).

Adopting this interpretation, the existence of this linear 
thermal filament indicates that 
the mean direction of the Cygnus A jet has remained 
approximately constant during the last 10 Myrs. 
While multiple hotspots represent recent changes 
in the jet direction,
the time-averaged jet direction has not deviated much
from the thermal X-ray jet-filament direction
during the age of Cygnus A.
Indeed, small asymmetries in the relatively dense 
thermal jet as it falls back near the 
cluster center may provide
an easy explanation for the occasional 
redirection and misalignment of the Cygnus A jet. 
Both thermal and non-thermal jets 
share the same initial direction. 
If the inertia of the non-thermal jet is small, 
as is likely, it could be easily deflected 
by thermal gas in the cluster core. 
But the sustained linear alignment of the 
thermal jet over time indicates that 
the basic structure of Cygnus A is axisymmetric. 
The gas density and (as yet undetected)  
thermal X-ray emission along the Cygnus A filament depends 
on the uncertain gas density profile 
near the center of the cluster gas 
prior to the FRII energy release. 
The density structure of the X-ray thermal filament 
shown in Figure 10, based on our assumed initial cluster gas 
profile, is only representative. 

The transverse cavity density profiles 
(Fig. 10 bottom) extend only to the 
cavity boundary defined by a gas density that is one third 
of the original cluster gas density at that radius.
Most of the mass inside the cavity is concentrated 
in the ``thermal sheath'' near 
its boundary. The origin 
of this sheath 
must be confirmed with calculations at higher resolution. 

\footnotetext[3]{
Radial post-cavity gas flows along the symmetry axis are analogous to 
the well-known reaction when a drop of water falls onto a perfectly
quiescent water surface. The impacted surface is briefly made concave
due to the momentum delivered by the falling drop, 
causing a splash. But immediately
afterwards, water below the concave surface flows 
toward the symmetry
axis as gravitational forces act to reestablish the original flat
surface.  But as this flow converges at the center of the 
depression, a small 
unidirectional jet of water squirts up far above the original
water surface.  The same effect occurs underneath a young 
approximately axisymmetric cluster cavity in which 
cluster gas flows down just outside the cavity walls, 
converges beneath the cavity and, without shocking, 
shoots radially up through the bubble, 
forming a supersonic thermal jet. 
The kinetic energy of the upflowing thermal jet comes from 
returning the potential energy it acquired 
during cavity formation.
Post-cavity thermal jets or upflows 
are often referred to as gas that is ``lifted'' or ``dragged'' 
out by the cavity, but it is a natural 
gasdynamical response to the cavitation.
Like its water analog, the thermal filament will 
eventually fall back 
to the cluster core on a dynamical time scale.
Gas further along the jet begins its fall back
at progressively later times.
}

\section{More Observational Verification Needed}

Our objective in this computation of radio lobe dynamics in 
Cygnus A and other similar FRII 
sources is to describe key morphological and kinematic components 
observed inside the lobes.
Consequently, we require a computational 
resolution comparable with those of current 
radio and X-ray observations.
Our computational accuracy is limited by several 
simplifying assumptions: 
(1) Our gasdynamics is non-relativistic.
Occasional, spatially limited appearances of 
near relativistic velocities with $(u/c)^2 \sim 0.2$ 
are tolerated 
since the leading order special relativistic 
corrections to the hydrodynamic equations are 
of order $(u/c)^2$ (van Odyck 2004).  
(2) As discussed in MG10, we assume $\gamma = 5/3$ 
for gas that is initially non-relativistic.
However, some gas inside the radio lobes rises to 
temperatures $\sim10^{10}$K for which $\gamma \rightarrow 4/3$,
introducing errors of $20$\%.
(3) Guided by observations of the Cygnus A hotspots and lobes, 
we assume that the magnetic field evolves passively with the 
gas flow and does not influence the overall lobe dynamics. 
Furthermore, we assume a simple toroidal magnetic field 
consistent with the radio polarization observed in 
the Cygnus A hotspots but not with that in the radio lobes.

However, it is unclear if 20-25\% errors due to 
these computational approximations can be tested 
with current radio and X-ray observations.
Cygnus A has received about 200ks of {\it Chandra} observations. 
Continued analysis of these observations, or still 
deeper observations, may help resolve 
the following questions of interest.

(1) Can the spatial and spectral variations of non-thermal 
(IC and SSC) X-ray emission be extracted from the thermal 
X-ray component in the radio lobe?
Is it possible to produce cylindrically deconvolved 
transverse profiles of thermal and non-thermal X-ray emissivities  
in Cygnus A or elsewhere? 
Is the non-thermal X-ray emissivity limb-brightened 
as expected?
How do the X-ray spectrum and surface brightness 
-- and the corresponding cylindrically deconvolved emissivity -- change 
across the  sharply defined radio lobe boundary? 
Does the X-ray spectrum vary along the boundary backflow? 
Is the X-ray cavity slightly wider than the radio lobe 
due to the hot ``thermal sheath'' seen in Figures 2 and 4? 
In MG10 we found deep, easily
observable X-ray cavities in Cygnus A 
when viewed only in thermal emission, 
but cavities are less apparent in the X-ray image in Figure 1.
Is the cavity by chance filled in with non-thermal X-ray emission?

(2) Is X-ray emission from the radial filament along the symmetry axis 
thermal?
Can its density be determined? 
Does this filament consist of cooler, relatively low entropy, 
metal rich gas as expected?
What is the nature of non-axisymmetric thermal X-ray emission within
about 30 kpc from the center (Fig 1 top)?

(3) What is the deconvolved (locally) 
axisymmetric variation of thermal and non-thermal X-ray  
emissivity within a few kpc of the Cygnus A (or other) hotspots? 
Can spatial variations in the hotspot wind be determined 
from radio or X-ray observations? 
Hotspot winds and the decay time of inactive hotspots 
may inform about the gas. field and CRs they contain. 

(4) Velocity shear in the boundary backflow is expected to 
amplify poloidal magnetic fields and deceleration is expected 
to amplify the toroidal component. 
Can radio and X-ray IC and SSC observations verify that the magnetic 
field increases along (or across) the boundary backflow?

(5) At 5GHz near the peak radio intensity in Cygnus A, 
the VLA FWHM is
nearly 1 kpc at Cygnus A (Wright \& Birkinshaw 2004),
comparable to the expected 
hotspot offset. 
Further radio observations of the hotspots with the new JVLA 
at 6-15 GHz with HPBW = 0.33-0.13 kpc may 
detect a radio offset. 
A detailed analysis of the Cygnus A X-ray hotspot 
may detect spatial and spectral differences between 
X-ray IC-CMB and SSC emission.

Finally, as pioneered by Carvalho et al. (2005) and 
Daly et al. (2010), 
whenever possible future radio and X-ray observations 
of bright axisymmetric FRII sources should be cylindrically deconvolved 
to determine emissivity profiles transverse 
to the lobe symmetry axis in physical units.

\section{Summary and Discussion}

We describe the axisymmetric gas and CR dynamical evolution 
of FRII radio sources similar to Cygnus A 
originating from a continuously active hotspot moving 
into an undisturbed cluster gas environment. 
The hotspot, regarded as the region where jet energy is isotropized, 
moves with constant 
velocity $v_{hs} = z_{hs}/t_{age} = 5870$ km s$^{-1}$ 
where $z_{hs} = 60$ kpc is the currently observed distance of 
the Cygnus A hotspots from the galaxy core and $t_{age} = 10$ Myr 
is the age of Cygnus A based on radio synchrotron aging observations. 
Most of the power $10^{46}$ erg s$^{-1}$
delivered by the hotspot to the expanding radio cavity 
and the surrounding cocoon shock 
is contained in cosmic rays (CRs) 
emerging behind the 
powerful reverse shock at the inner boundary of 
the hotspot where the jet impacts.
We assume that the contents of the evolving hotspot 
are continuously renewed to resemble 
those currently observed in Cygnus A: 
CR energy, a passive toroidal magnetic field and 
a small mass of hot gas. 

Computations of FRII flows 
including only gas pressure gradients and gravity 
match several, but not all important attributes of 
observed FRIIs.
The cavity volume and cocoon shock are similar to those 
observed in Cygnus A at time 10 Myrs (MG10). 
Gas and CRs flow rapidly from the hotspot source, 
but are forbidden to flow upstream directly into the 
oncoming jet.
This boundary condition 
helps re-direct CRs and gas in the hotspot outflow 
back along the boundary of the radio cavity, 
forming a boundary backflow.
To determine the CR age in the radio cavity,
we compute the product of the instantaneous gas density and time in
the hotspot, $\tau = \rho t$, and follow its subsequent flow 
inside the radio lobe.
In straightforward FRII computations 
with gas pressure gradients and gravity, 
Kelvin-Helmholtz (KH) shear instabilities 
scramble CR ages in the interior of the radio cavity 
and wrinkle the cavity boundary,
neither of which are observed.

Therefore it is necessary to seek solutions in which the KH features 
are damped, either by magnetic fields or viscosity.
Magnetic KH damping might be possible if the 
thermal gas density in the radio cavities 
is much lower than we expect. 
But we explore here the possibility that 
undesired KH features are damped by viscous forces.
Viscous and diffusive 
transport properties in a weakly magnetic radio cavity 
containing gas and relativistic CRs are poorly understood, 
particularly since  
the mean free path for Coulomb scattering is vastly 
larger than the Larmor radius. 
The viscosity we invoke may be turbulent in origin. 
Regardless of how the KH instability is damped, 
the boundary backflow is expected to shear and decelerate. 
Along the boundary backflow 
poloidal fields will be amplified by shear and 
toroidal fields by compression. 

The flow inside the radio cavity 
becomes ordered in a remarkable way when the 
KH instability is damped with a small uniform viscosity.
As in our non-viscous computations, 
the radio synchrotron emission is confined 
in a boundary backflow.
But damping the KH instability allows 
CR ages to remain spatially ordered 
long after they leave the hotspot 
and surface KH features largely disappear.
Viscosity provides the same excellent 
improvement for FRII sources as it does for the 
Fermi bubbles in the Milky Way 
(Guo \& Mathews 2012; Guo et al. 2012).
With viscosity included 
the CR boundary backflow persists throughout the 
length of the cavity 
as commonly observed for FRII sources (e.g. Alexander \& Leahy 1987).
The radio synchrotron emissivity is very strongly 
limb-brightened at the radio lobe boundary, 
similar to recent detailed radio observations
of Cygnus A and other FRII sources 
(Carvalho et al. 2005; Daly et al. 2010).

We also explore the evolution of passive toroidal magnetic fields 
consistent with radio polarization 
morphology observed in the Cygnus A hotspots.
CRs and magnetic field flow out from the hotspot, 
filling the cavity.
Our calculation may be the first to establish an 
approximate 
relationship between the magnitude of the hotspot field 
and the field observed in the much larger radio cavity.
While the field orientation in the Cygnus A hotspot is 
mostly toroidal, radio polarization observations 
reveal non-toroidal field components downstream inside 
the radio cavity.
An idealized perfectly toroidal field expelled from an ideally 
axisymmetric hotspot remains toroidal after 
flowing into the radio lobe. 
Toroidal fields cannot suppress KH instabilities, 
do not experience laminar shear amplification, 
and may weaken or suppress viscous momentum diffusion 
in the $r$-direction across the boundary backflow.
The toroidal field we consider 
intensifies by compression along the 
decelerating backflow. 
If the hotspot contains additional poloidal
field components, they would be shear-amplified 
in the radio cavity backflow, but the overall 
amplification may not be as large as that of 
the purely toroidal fields we consider.
Radio lobe fields may also be isotropized by the
turbulent viscosity we require to smooth the lobe
kinematics.
Nevertheless, we find that a purely toroidal hotspot 
field of $B_{hs} \sim 200\mu$G, 
as suggested by the SSC models of Stawarz et al. (2007),  
remains much too large 
(even exceeding $B_{hs}$) after evolving into the 
decelerating radio lobe region. 
We speculate that the field in the hotspot may 
have been overestimated due to nearby downstream 
regions with larger fields 
or to inhomogeneities inside the complex post-shock hotspot
interior,
but it is unclear if these alone can fully explain
the hotspot-lobe field discrepancy. 

CRs and hot gas flowing in the hotspot wind 
push aside nearby gas in the radio lobe. 
In particular, as the CR-dominated hotspot wind 
decelerates toward 
the dense wall of shocked cluster gas 
directly ahead of the hotspot, 
the magnetic field in the wind is amplified. 
This has the effect of enhancing the radio 
synchrotron emission from this region, 
producing a small arc-like radio-bright region 
offset 1-2 kpc ahead of the hotspot. 
The radio surface brightness in this arc exceeds that 
from the hotspot. 
Nonthermal IC-CMB X-ray emission probably peaks
near the hotspot shock, while 
SSC X-ray emission peaks 
closer to the offset radio synchrotron peak, 
reducing the overall radio-X-ray offset. 

While the position and velocity of the hotspot 
are fully prescribed in our calculation, we expect that 
the reverse shock in an unconstrained hotspot is 
naturally located by 
subsonic hydrodynamic communication with 
gas behind the bow shock. 
High resolution computations will be required 
to accurately determine the detailed structure of this 
complex hotspot region.
The properties of the hotspot wind 
required to match the hotspot offset morphology 
may constrain the particle and field 
contents of both the hotspot and jet.

Our computation also reproduces the faint, presumably 
thermal X-ray radial filament visible in 
X-ray images of Cygnus A.
This narrow filament is an expected gasdynamical reaction 
to the formation of an axisymmetric low density 
cavity in the cluster gas.

Our computations are approximate in several ways: 

\noindent$\bullet$ We do not 
treat with precision the spatially limited appearance of 
mildly relativistic velocities 
nor the 
transition of thermal cluster gas to mildly relativistic 
temperatures. 

\noindent$\bullet$ We overlook 
a possible inconsistency inside the radio cavity 
in which viscous momentum transport 
occurs perpendicular to our assumed 
toroidal magnetic field passively evolving 
from the hotspot. 
However, radio polarization in the Cygnus A 
cavity indicates that the radio lobe field is not purely toroidal.

\section{Conclusions}

Our gas dynamical calculation is successful in matching 
many detailed observations of Cygnus A and other FRII 
sources:

\noindent$\bullet$ We describe a ``boundary backflow'' from which most 
of the strongly limb-brightened radio synchrotron emission occurs 
in agreement with FRII observations.

\noindent$\bullet$ The trend in ages of radio synchrotron electrons 
along and transverse to the boundary backflow 
are monotonically organized, resembling observations. 

\noindent$\bullet$ To achieve 
this smooth variation of apparent synchrotron 
ages and smooth radio lobe boundaries, it is necessary to 
damp Kelvin-Helmholtz shear instabilities. This can be done 
with a small viscosity of uncertain origin, which we propose 
here, or possibly with magnetic tension provided 
the plasma density inside the radio lobe is considerably smaller 
than expected. 

\noindent$\bullet$ Our approximate calculation of 
toroidal magnetic fields passively evolving 
from the hotspot into the radio lobe allows us to relate the 
observed field strength in these two regions. 
If radio cavity fields evolve from toroidal fields
estimated in the Cygnus A hotspots,
the fields 
in the radio-emitting backflow are
about 10-20 times larger than those observed, $15-20\mu$G.
The origin of this disparity is unclear.
%
%
Perhaps the field 
in the physically complex hotspots is lower than expected. 
The CR energy density inside our computed hotspot,  
required to inflate the Cygnus A cavity, is about 
ten times larger than the total CR electron energy density 
previously estimated, suggesting an additional
hotspot component of non-radiating particles.

\noindent$\bullet$ Low surface brightness X-ray emission 
in Cygnus A along 
the symmetry axis of each radio lobe can be understood as 
a jet or filament of thermal gas flowing from the cluster core, 
an expected hydrodynamic feature that accompanies 
all rapidly formed axisymmetric cavities in cluster gas. 
We speculate that small asymmetries in 
the relatively dense innermost part of 
this thermal jet, perhaps as it falls back, are responsible 
for sudden changes in the direction of the non-thermal 
jet in Cygnus A, particularly since the non-thermal jet is likely 
to have a rather small inertia.

\noindent$\bullet$ When viewed in projection, 
our computed FRII flows predict two spatially distinct regions 
of enhanced non-thermal emission associated with the hotspot 
and its wind.
Enhanced CRs directly behind the hotspot shock 
cause this region to emit nonthermal IC-CMB X-rays.
Radio synchrotron emission is strongest in a 
second, more extended arc-shaped region 1-2 kpc 
ahead in the decelerating hotspot wind 
where the magnetic field is increased by compression.
Nonthermal SSC X-ray emission, expected to 
be somewhat more luminous than IC-CMB X-ray emission,
peaks in the hotspot wind 
near the radio synchrotron offset. 
Nonthermal X-ray emission is expected throughout the
hotspot-arc region.

\noindent$\bullet$ A narrow layer of very hot
gas appears to backflow along the cavity 
wall just outside the radio cavity, 
causing the X-ray cavity to appear slightly larger 
than the radio lobe. 
This ``thermal sheath'', marginally resolved in our calculation, 
may contain most of the thermal gas inside the X-ray cavity.

\section{Acknowledgments}
We have benefited significantly from helpful advice suggested 
by the referee. 
Studies of feedback gasdynamics in hot intracluster gas 
at UC Santa Cruz are supported by NSF and NASA grants
for which we are very grateful.


\vskip.2in
\centerline{APPENDIX}
\vskip.2in

\centerline{Viscous Terms}
\vskip.2in
In cylindrical coordinates the $r$ and $z$ components of 
${\bf \nabla \cdot \Pi}$ are 
\begin{displaymath}
\mu \left[ {\partial^2 u_r \over \partial r^2}
+{\partial^2 u_r \over \partial z^2}
+ {\partial \over \partial r}{\left( {u_r \over r} \right)}
\right]
+ {\mu \over 3} {\partial ({\bf \nabla \cdot u})
\over \partial r}
\end{displaymath}
and
\begin{displaymath}
\mu \left[
  {\partial^2 u_z \over \partial r^2}
+ {\partial^2 u_z \over \partial z^2}
+ 2 {\partial u_z \over \partial (r^2)}
\right]
+ {\mu \over 3} {\partial ({\bf \nabla \cdot u}) \over \partial z}
\end{displaymath}
respectively where the viscosity $\mu$ is regarded as constant.
The viscous term ${\bf \Pi:\nabla u}$ in the internal energy equation is 
\begin{displaymath}
 2 \mu \left[
\left({\partial u_r \over \partial r}\right)^2
+ \left({u_r \over r}\right)^2
+\left({\partial u_z \over \partial z}\right)^2
+ {1 \over 2}
\left({\partial u_z \over \partial r}
   + {\partial u_r \over \partial z}\right)^2 \right]\\
\end{displaymath}
\begin{displaymath}
- {2 \over 3}\mu ({\bf \nabla \cdot u})^2.
\end{displaymath}


                                                                                   
\clearpage

\begin{figure}
\centering
\vskip0.9cm
\includegraphics[width=3.7in,angle=0]
{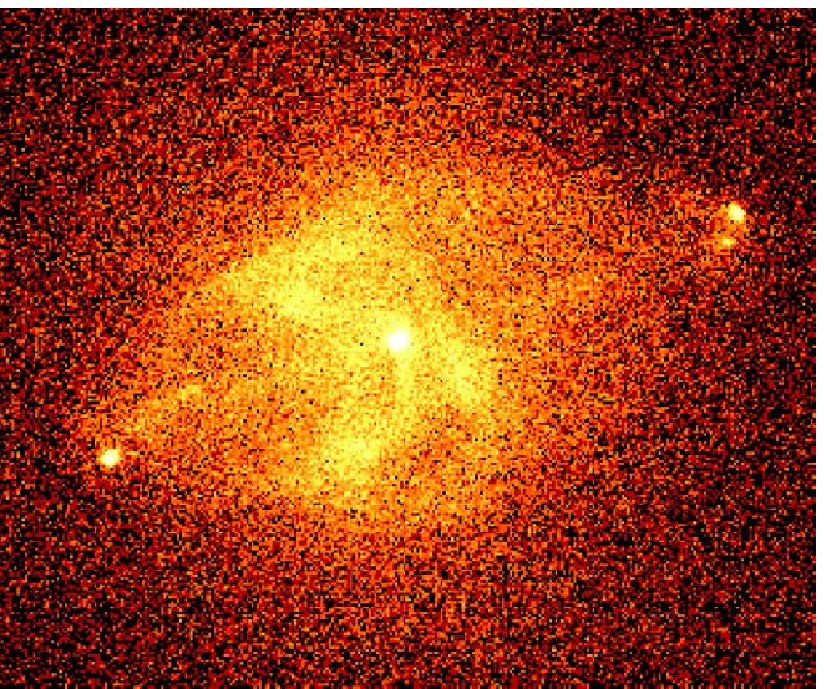}
\hspace*{-.6in}
\includegraphics[width=4.in,scale=0.8,angle=0]
{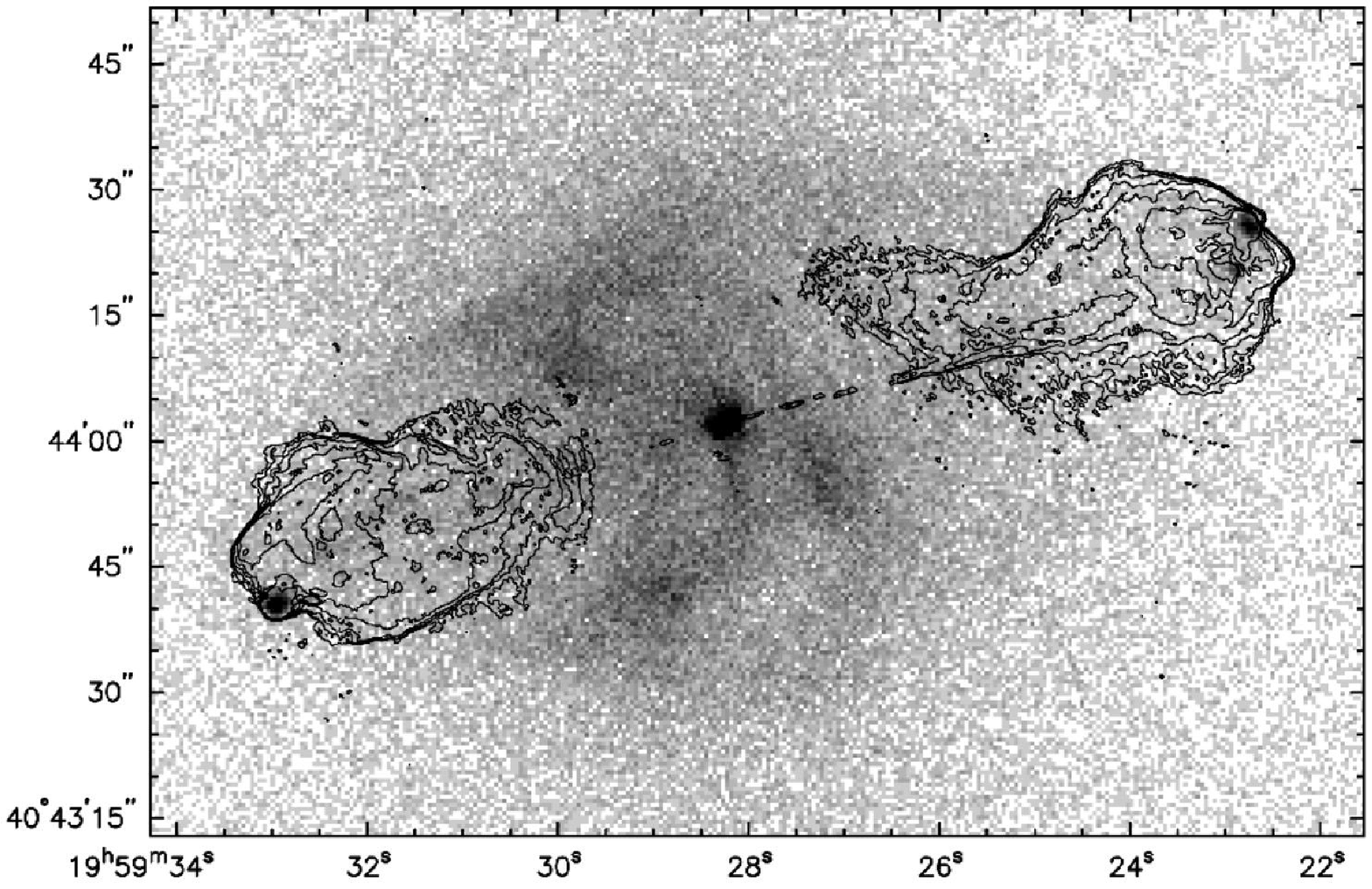}
\includegraphics[width=4.in,scale=0.8,angle=0]
{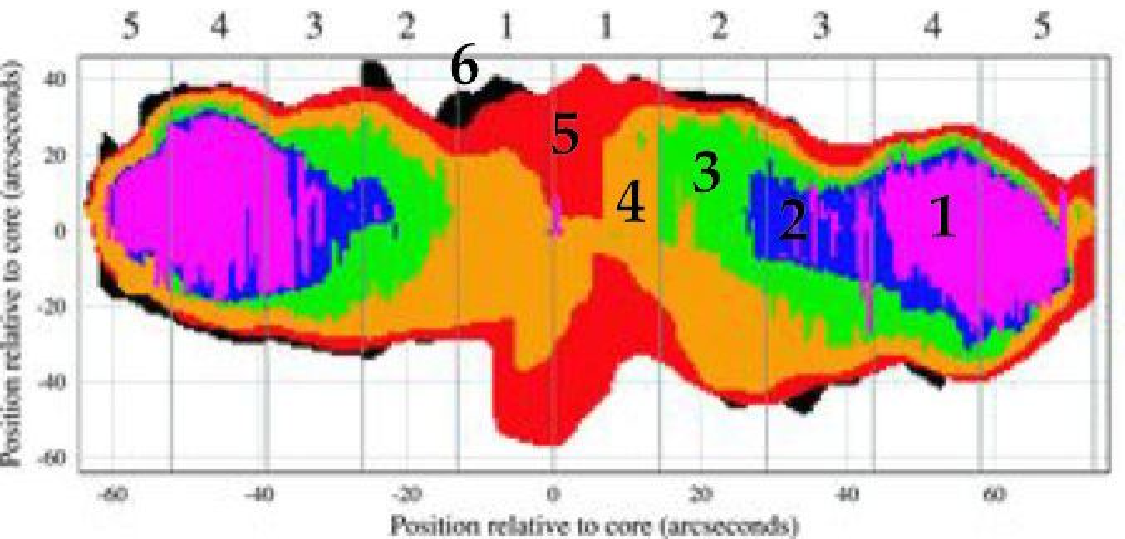}
\caption{
{\it Top:} Chandra image of Cygnus A is 150 kpc                                  
wide (1'' = 1 kpc).                                                             
Two oppositely-directed jets create a football-shaped                           
shock wave enclosing a cocoon of shocked gas.
{\it Center:} Same image with VLA contours at 5GHz.                              
(Wilson et al. 2006)
{\it Bottom:} Rotated Cygnus A at six radio frequencies                          
(Steenbrugge et al. 2010): (1) 15 GHz, 
(2) 8 GHz, (3) 5 GHz, (4) 1345 MHz,
(5) 327 MHz, and (6) 151 MHz. 
}
\end{figure}

\clearpage

\begin{figure}
\centering
\includegraphics[width=4.0in,scale=0.45,angle=0]
{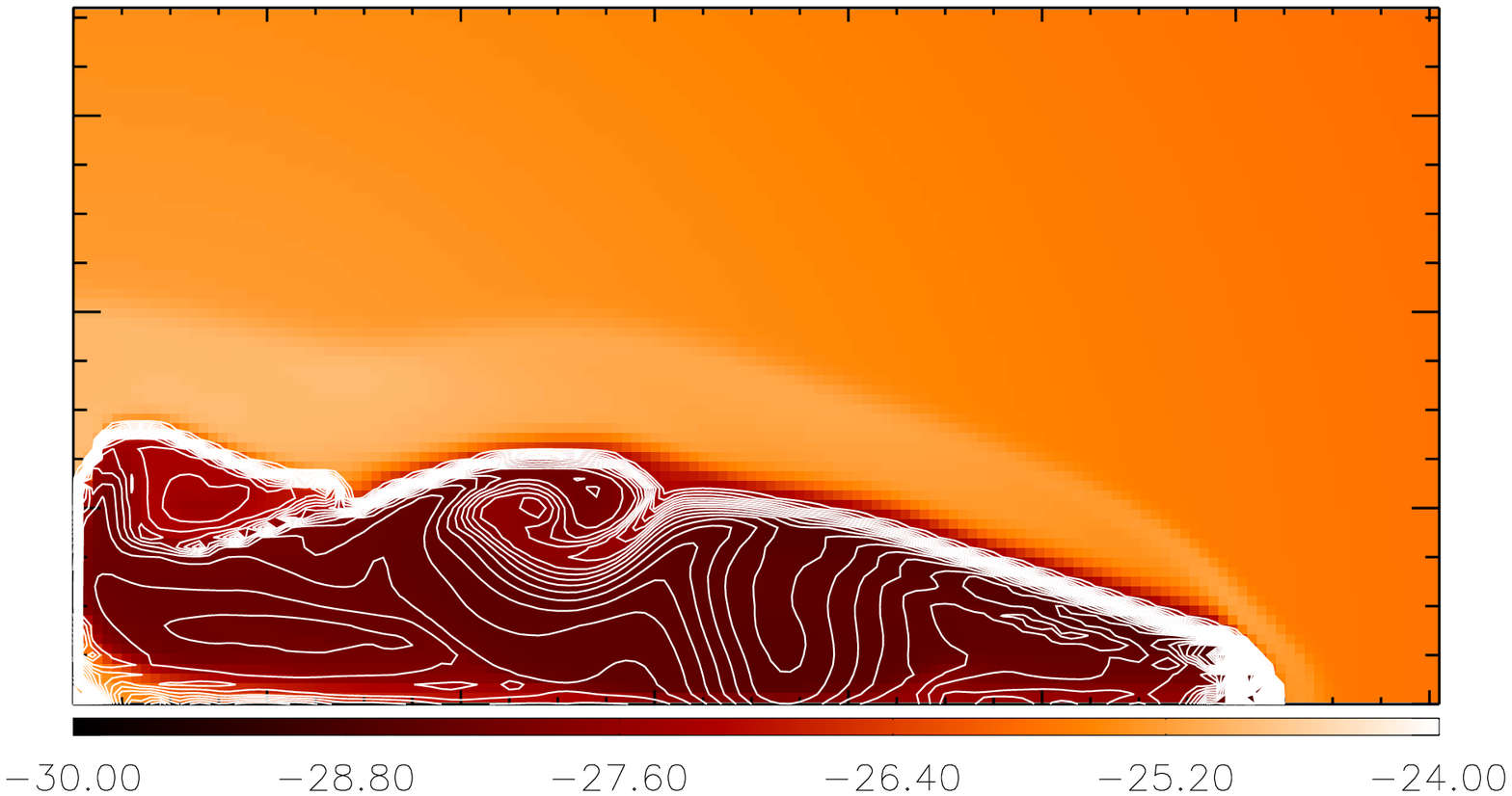}
\vskip.2in
\hspace*{-.3in}
\includegraphics[width=3.5in,scale=0.8,angle=0]
{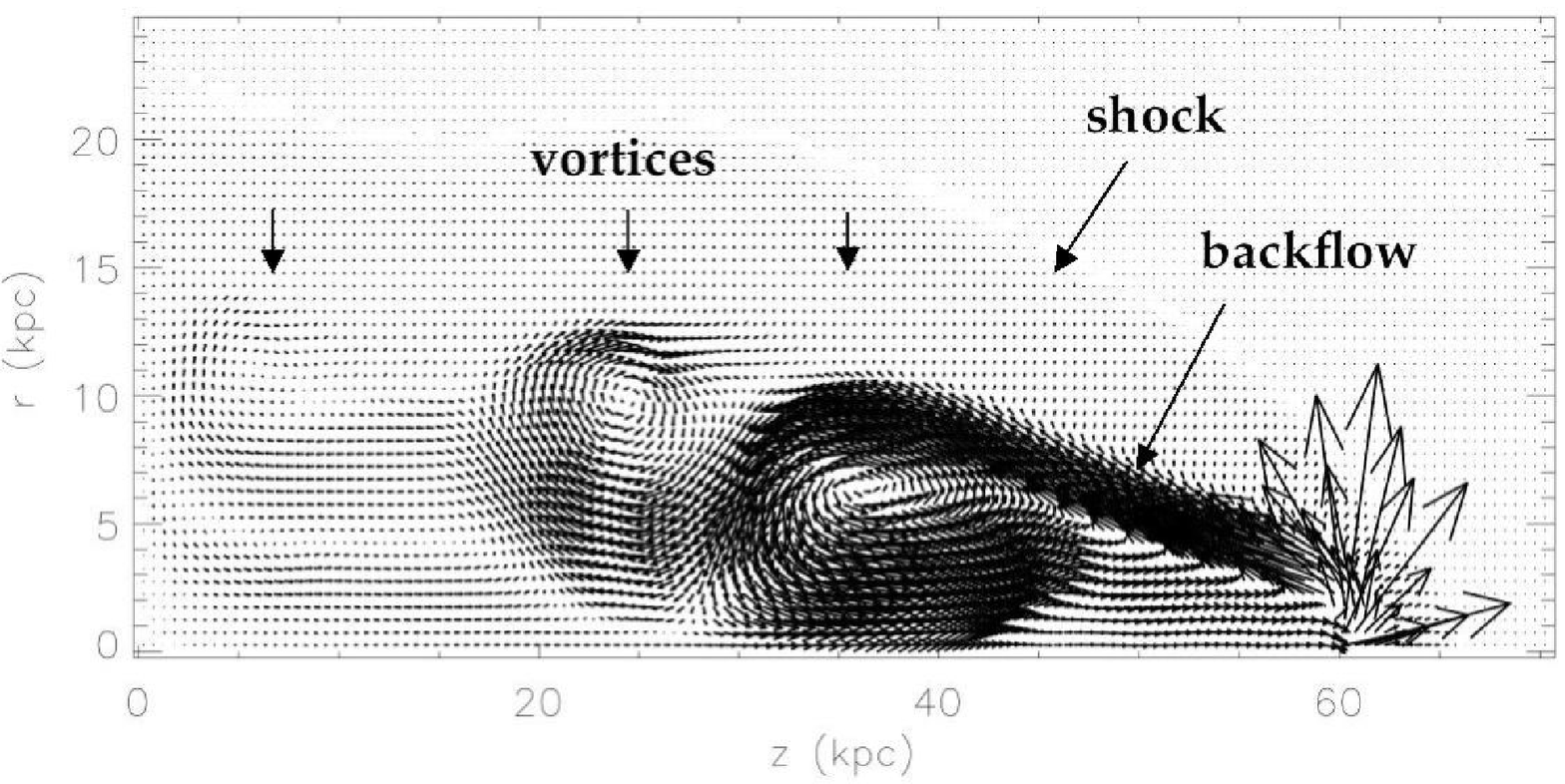}
\vskip.05in
\hspace{-.3in}
\includegraphics[width=3.7in,scale=0.8,angle=0]
{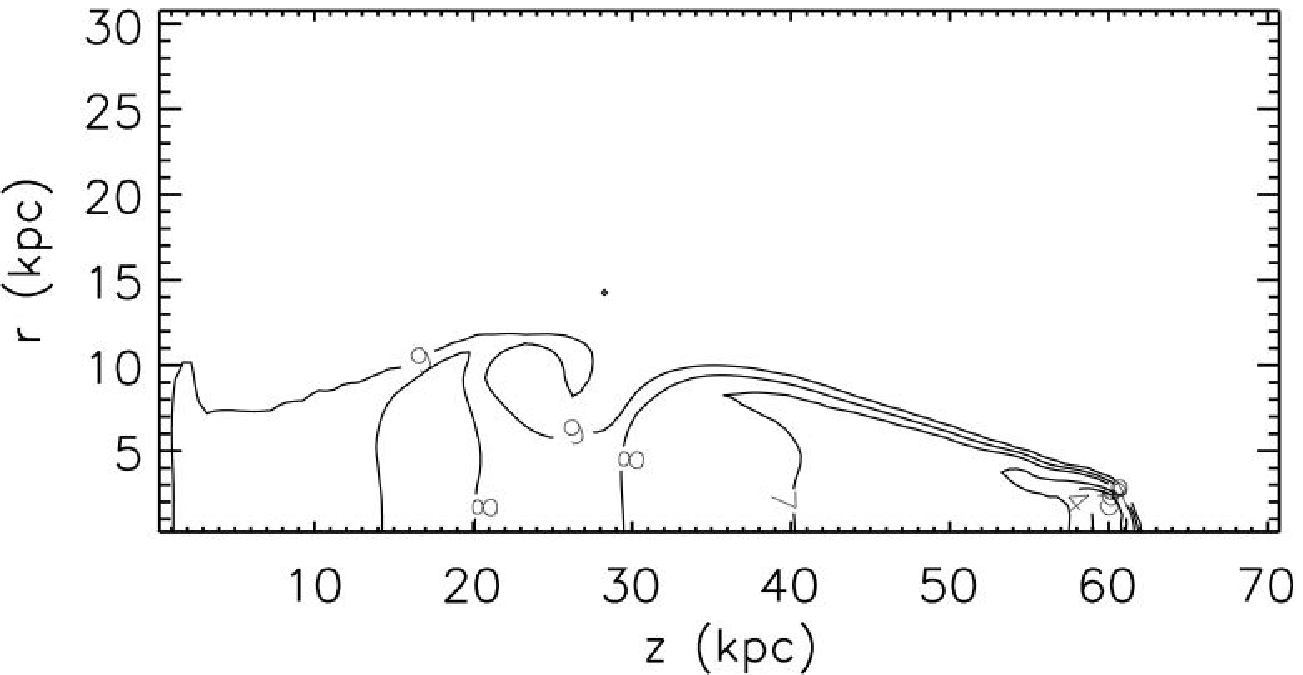}
\caption{
Quadrant-cut of invisid flow in Cygnus A at time 10 Myrs.
{\it Top:} Hot gas density $\log \rho(z,r)$ with cosmic 
ray energy density $e_c(z,r)$ shown with white contours.
{\it Center:} Total flow velocity ${\bf u}(r,z)$ shown with 
many overlapping arrows. 
{\it Bottom:} Contours show the mean emission-weighted 
line of sight CR synchrotron age $\langle t_{age} \rangle(z,r)$
in Myrs inside the radio lobe. 
The three most extended contours show age contours 
of 9, 8 and 7 Myrs. 
Smaller contours are at ages 4 and 5 Myrs.
The $z$-axis (horizontal) and $r$-axis (vertical) are 
shown in kpc in the lower two panels. 
Large tick marks in the top image are 
separated by 10 kpc. 
}
\end{figure}

\clearpage

\begin{figure}
\centering
\includegraphics[width=4.in,scale=0.45,angle=270]
{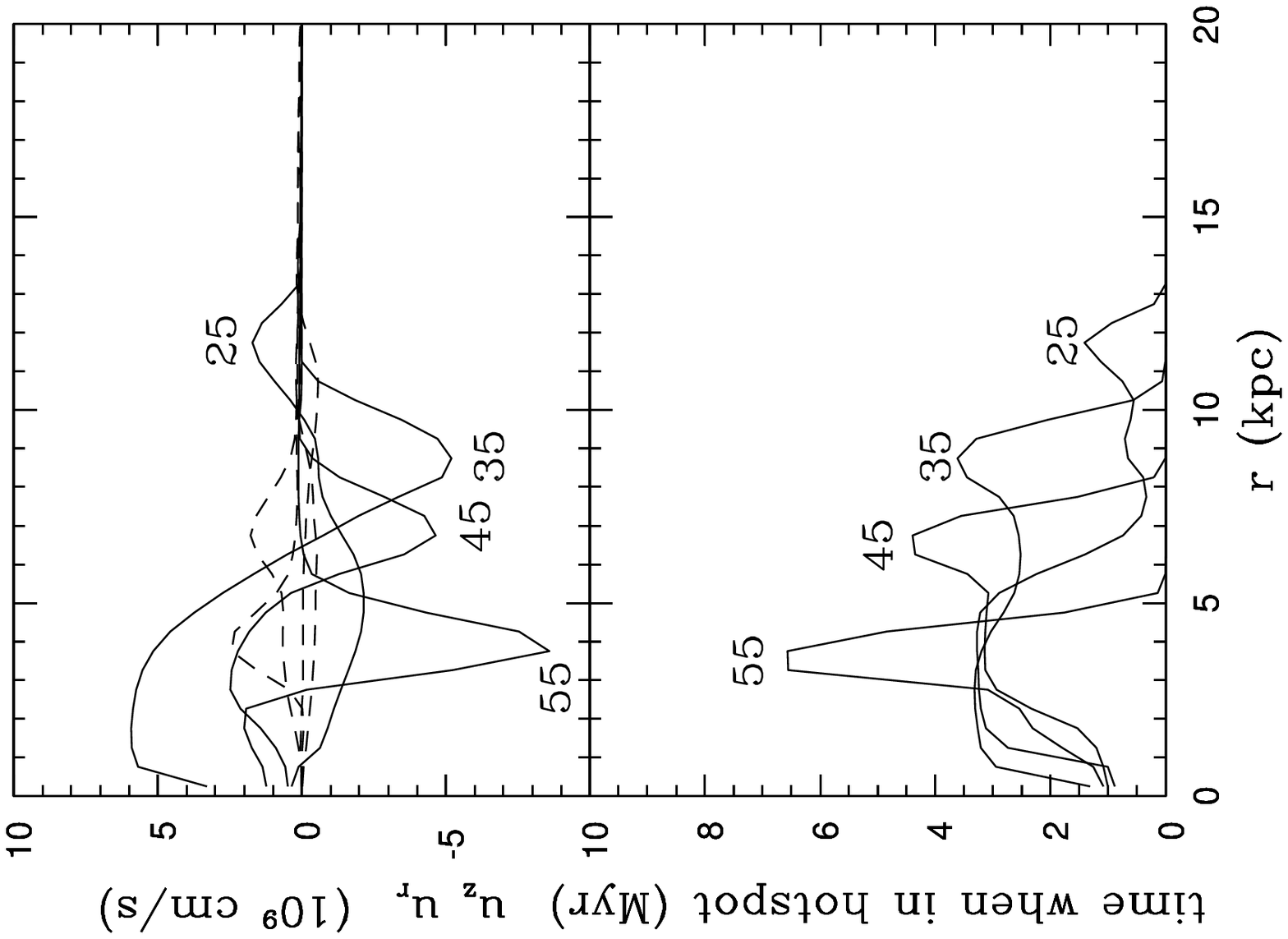}
\caption{
Profiles for non-viscous computation.
{\it Top:} Flow velocities $u_z$ ({\it solid lines}) and $u_r$
({\it dashed lines}) 
at four distances from the Cygnus A center: $z = 25$, 35, 45 and 55
kpc. At small $z$ the profiles for $u_r$ resemble lower amplitude 
positive reflections of the $u_z$ profile.
{\it Bottom:} Time in Myrs when gas left the hotspot shown 
at four distances from the Cygnus A center: $z = 25$, 35, 45 and 55 kpc.
}
\end{figure}

\clearpage

\begin{figure}
\centering
\includegraphics[width=4.in,scale=0.45,angle=0]
{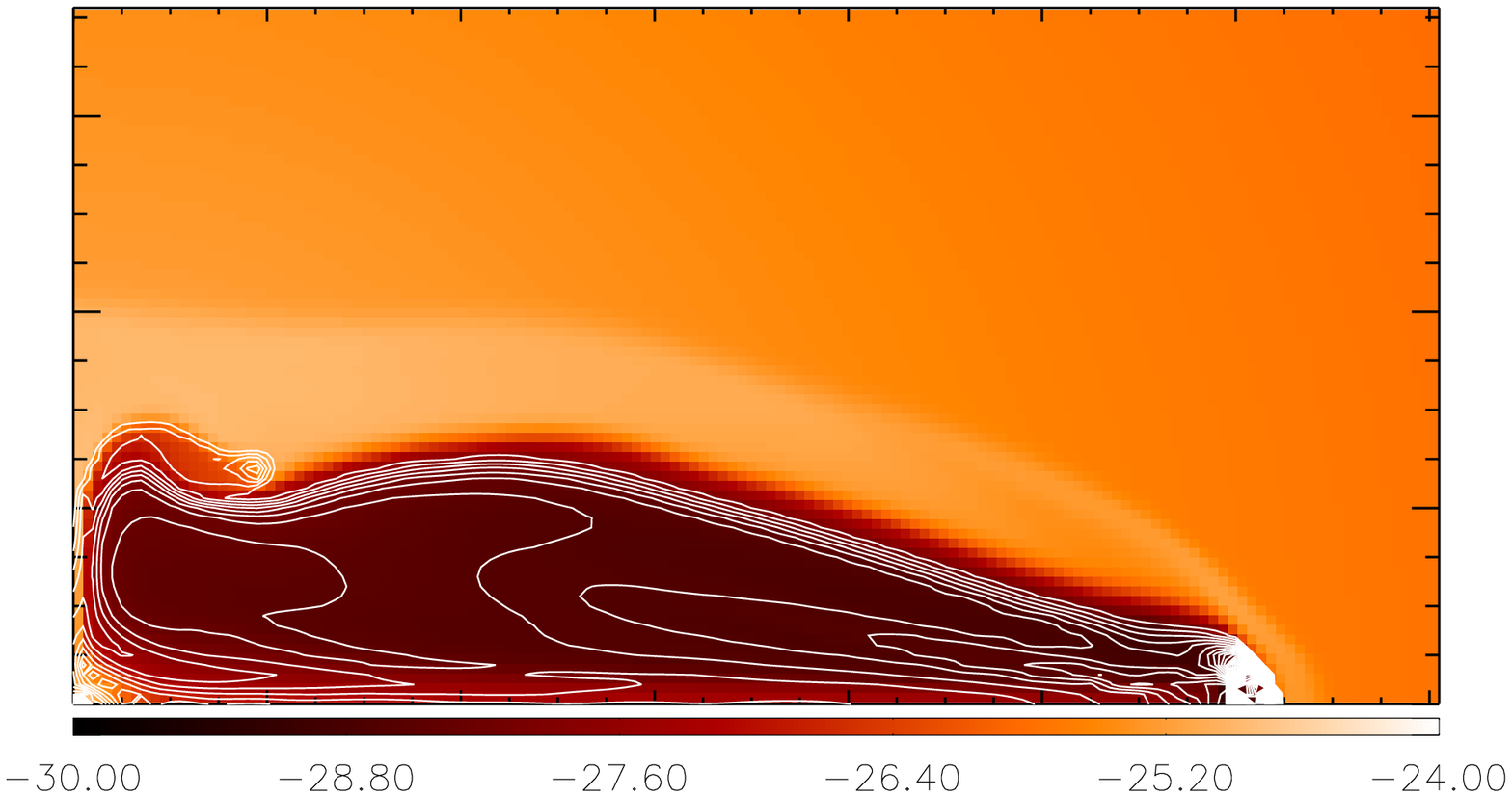}
\vskip-.3in
\hspace*{-.3in}
\includegraphics[width=4.in,scale=0.8,angle=0]
{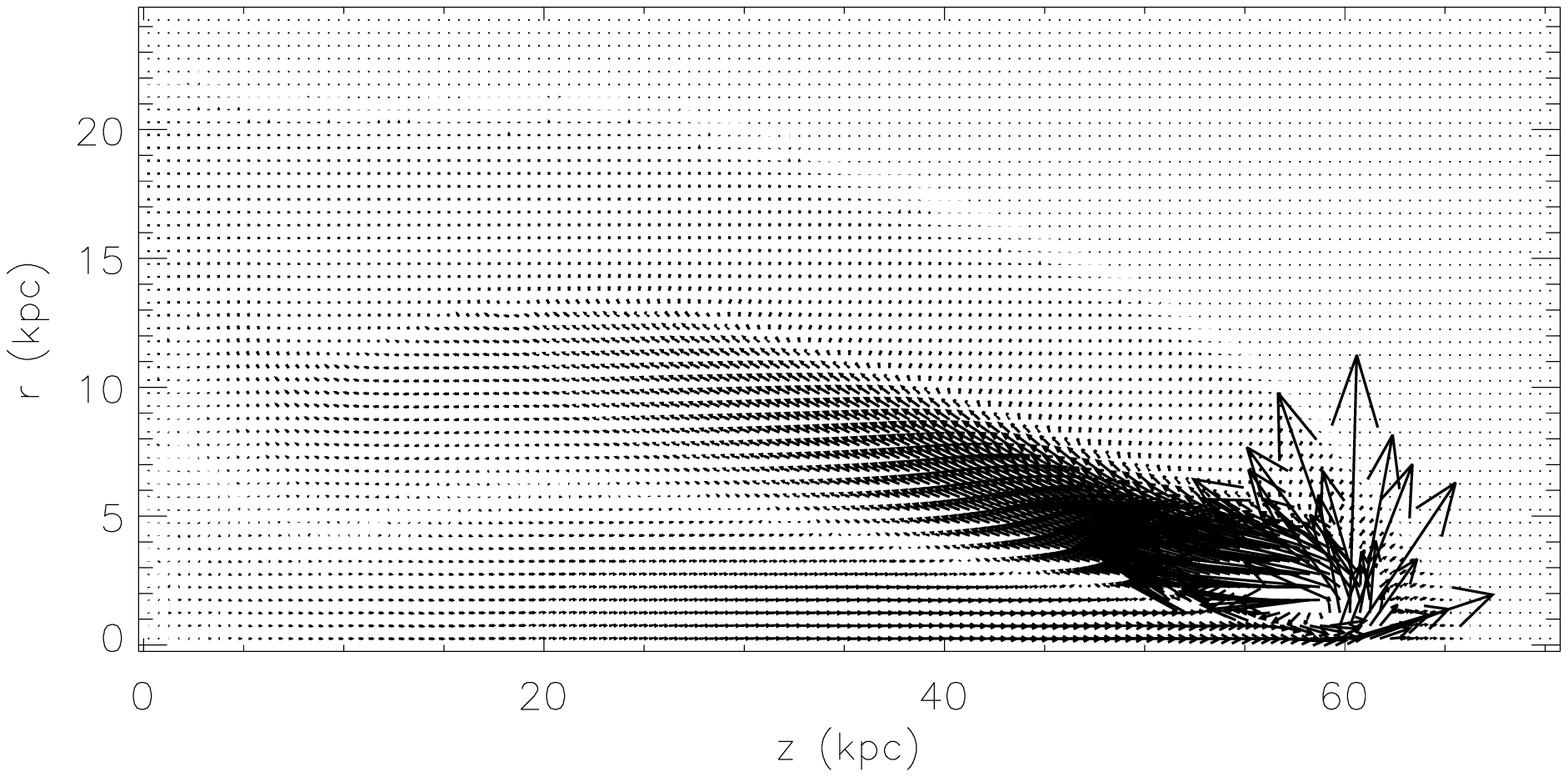}
\vskip-.10in
\hspace{-.3in}
\includegraphics[width=3.7in,scale=0.8,angle=0]
{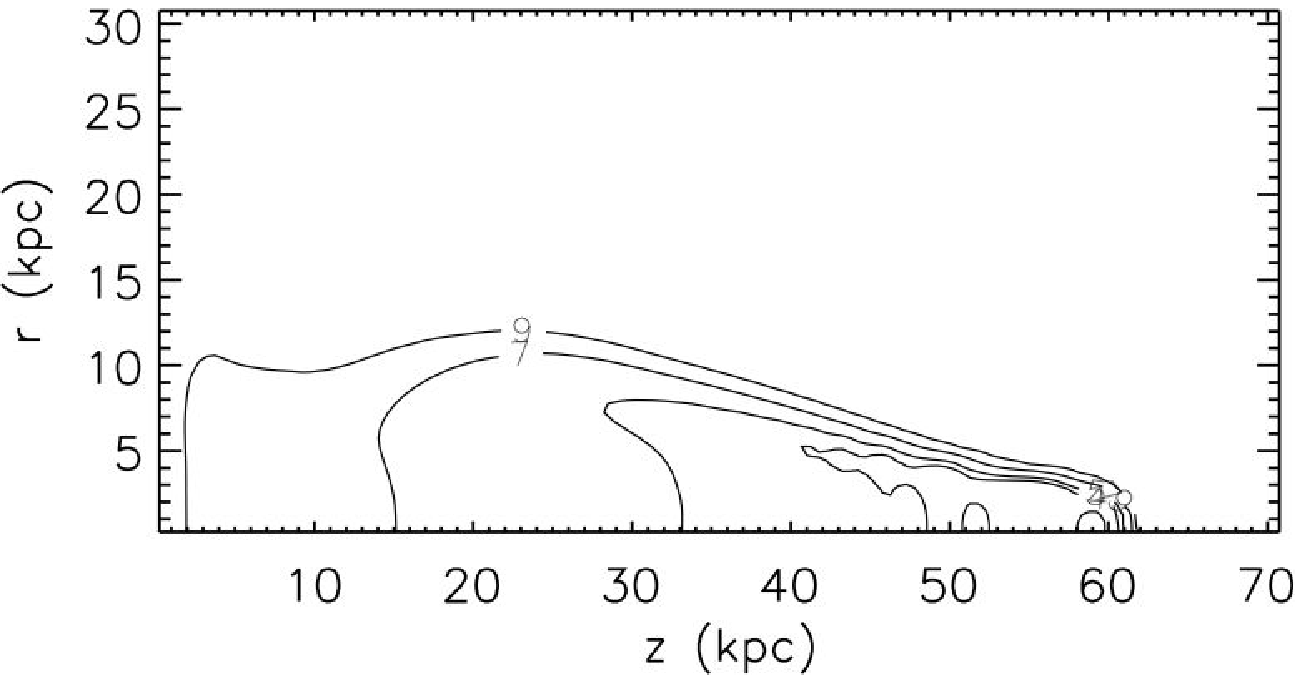}
\caption{
Quadrant-cut of flow in Cygnus A with viscosity 
$\mu = 30$ cm$^{-1}$ s$^{-1}$ at time 10 Myrs.
{\it Top:} Hot gas density $\log \rho(z,r)$ with cosmic                                       
ray energy density $e_c(z,r)$ shown with white contours. 
{\it Center:} Total flow velocity ${\bf u}(r,z)$. 
{\it Bottom:} Contours show the mean emission-weighted
line of sight CR synchrotron age $\langle t_{age} \rangle(z,r)$
in Myrs inside the radio lobe.
The two most extended contours at age 9 and 7 Myrs are 
clearly labeled. 
The contour that crosses the $z$-axis at 33 kpc 
is the CR age at 5 Myrs 
and the innermost contour is for 4 Myrs.
The $z$-axis (horizontal) and $r$-axis (vertical) are
shown in kpc in the central and lower panels.
}
\end{figure}

\clearpage

\begin{figure}
\centering
\includegraphics[width=4.in,scale=0.45,angle=270]
{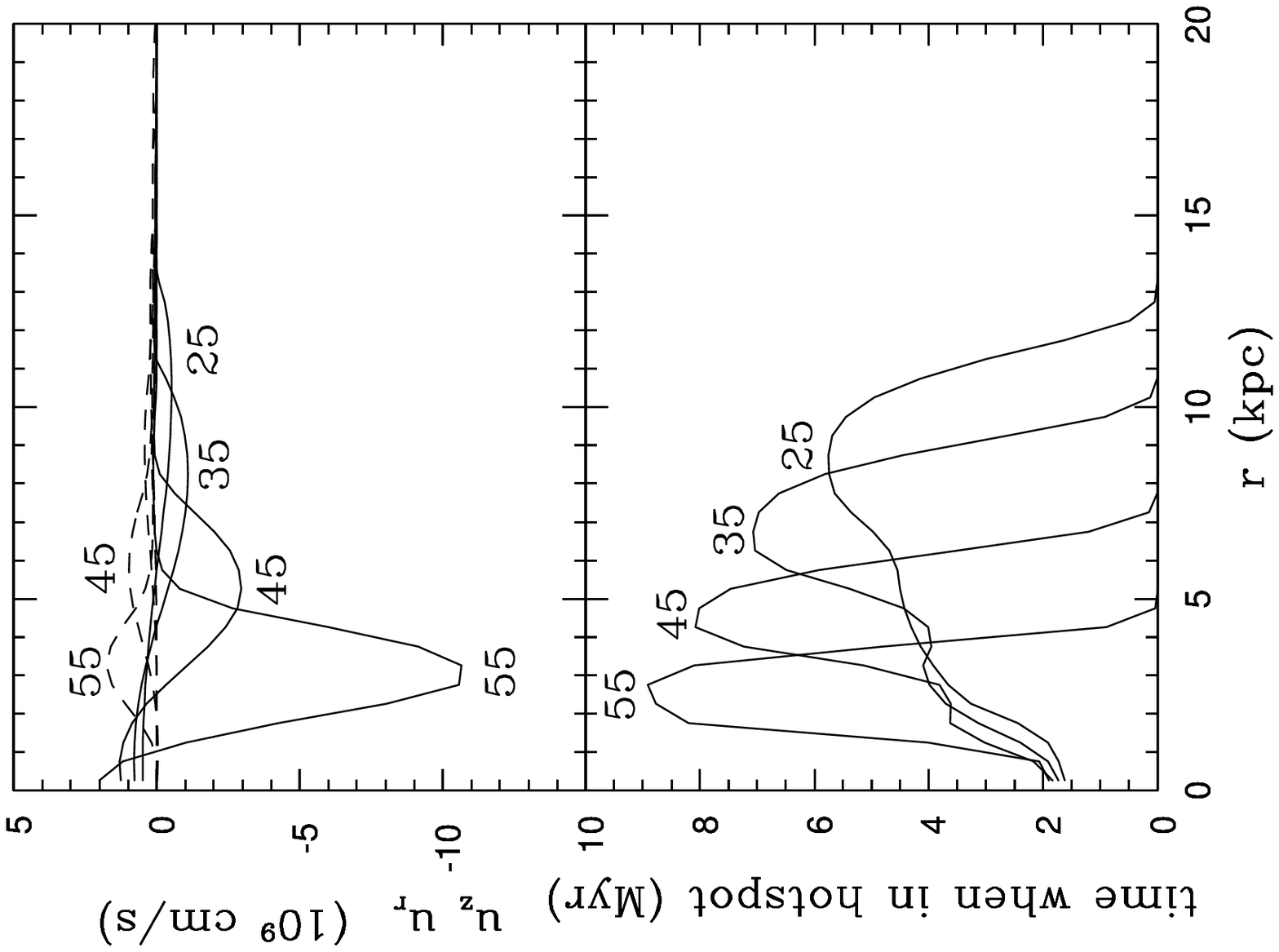}
\caption{
Profiles for viscous computation.
{\it Top:} Flow velocities $u_z$ ({\it solid lines}) and $u_r$
({\it dashed lines})
at four distances from the Cygnus A center: $z = 25$, 35, 45 and 55
kpc. At small $z$ the profiles for $u_r$ resemble lower amplitude
positive reflections of the $u_z$ profile.
{\it Bottom:} Time in Myrs when gas left the hotspot shown
at four distances from the Cygnus A center: $z = 25$, 35, 45 and 55
kpc.
}
\end{figure}

\clearpage

\begin{figure}
\centering
\includegraphics[width=4.in,scale=0.45,angle=0]
{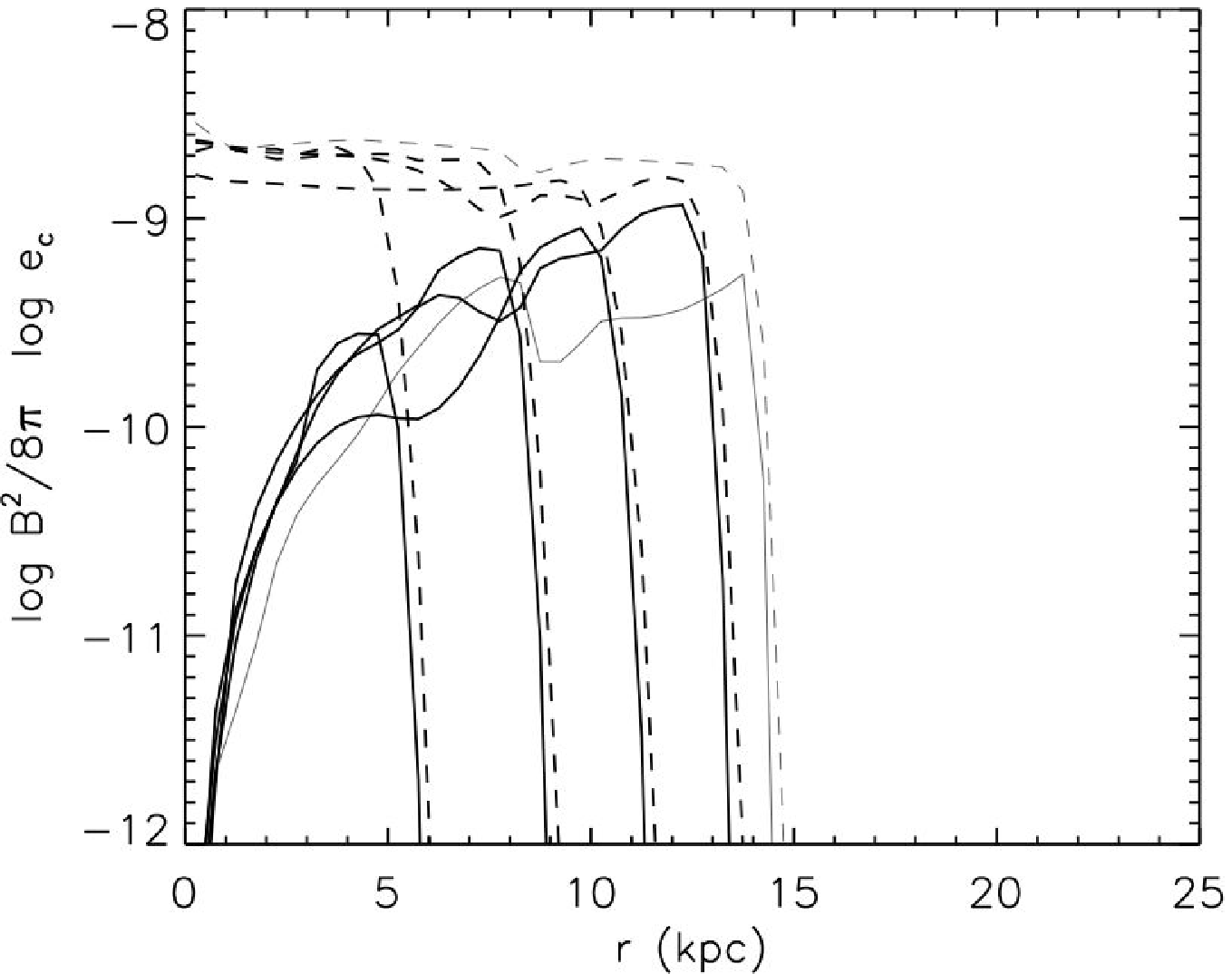}
\vskip-.2in
\includegraphics[width=4.in,scale=0.45,angle=0]
{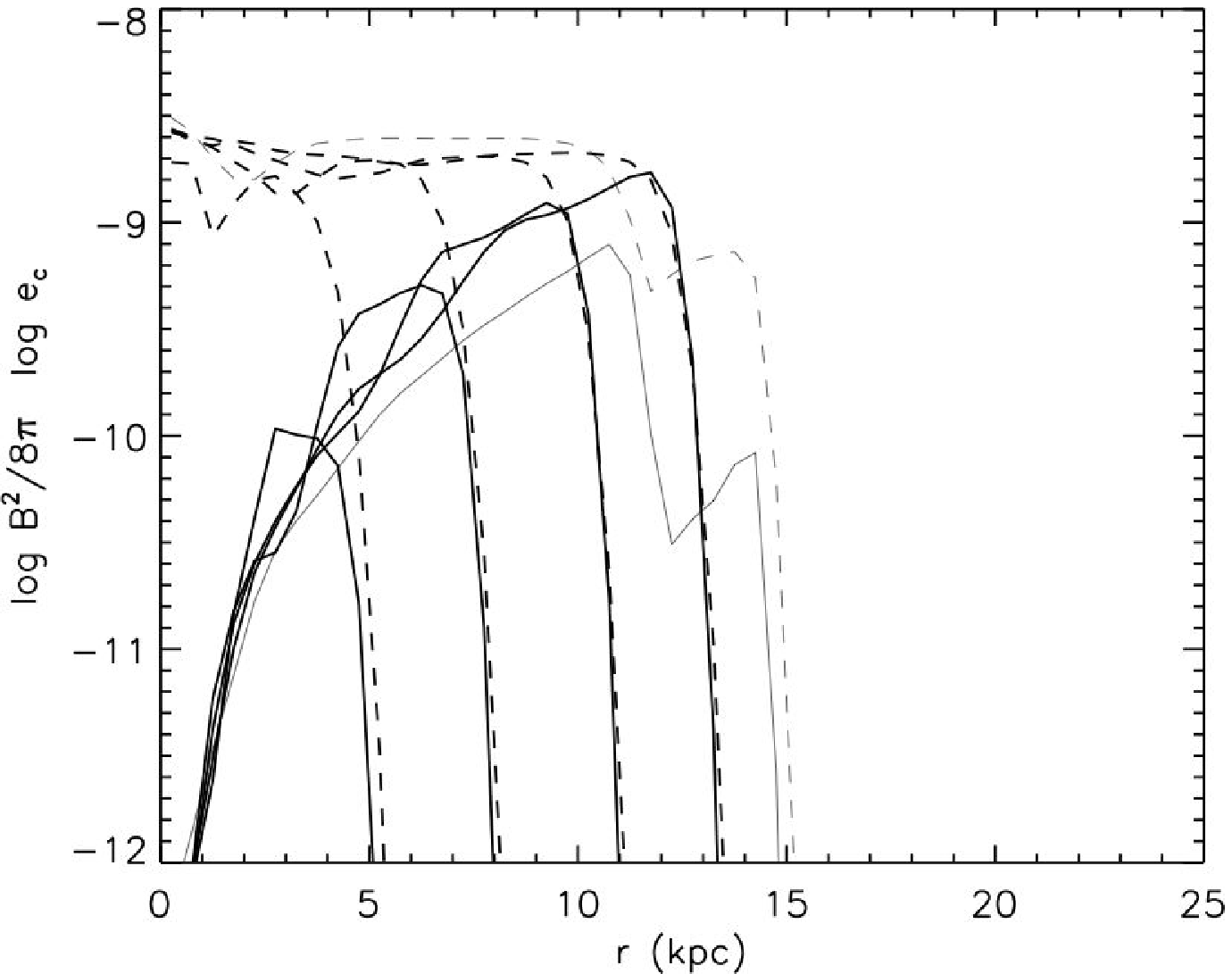}
\caption{
Transverse contours of CR energy density ({\it dashed lines}) 
and magnetic field energy density ({\it solid lines}) 
from left to right 
at $z = 55$ 45, 35, 25 ({\it heavy lines}) and 5 ({\it light lines})
kpc. Units are erg cm$^{-3}$.
{\it Upper panel:} non-viscous flow 
{\it Lower panel:} viscous flow.
}
\end{figure}

\clearpage

\begin{figure}
\centering
\includegraphics[width=4.in,scale=0.45,angle=0]
{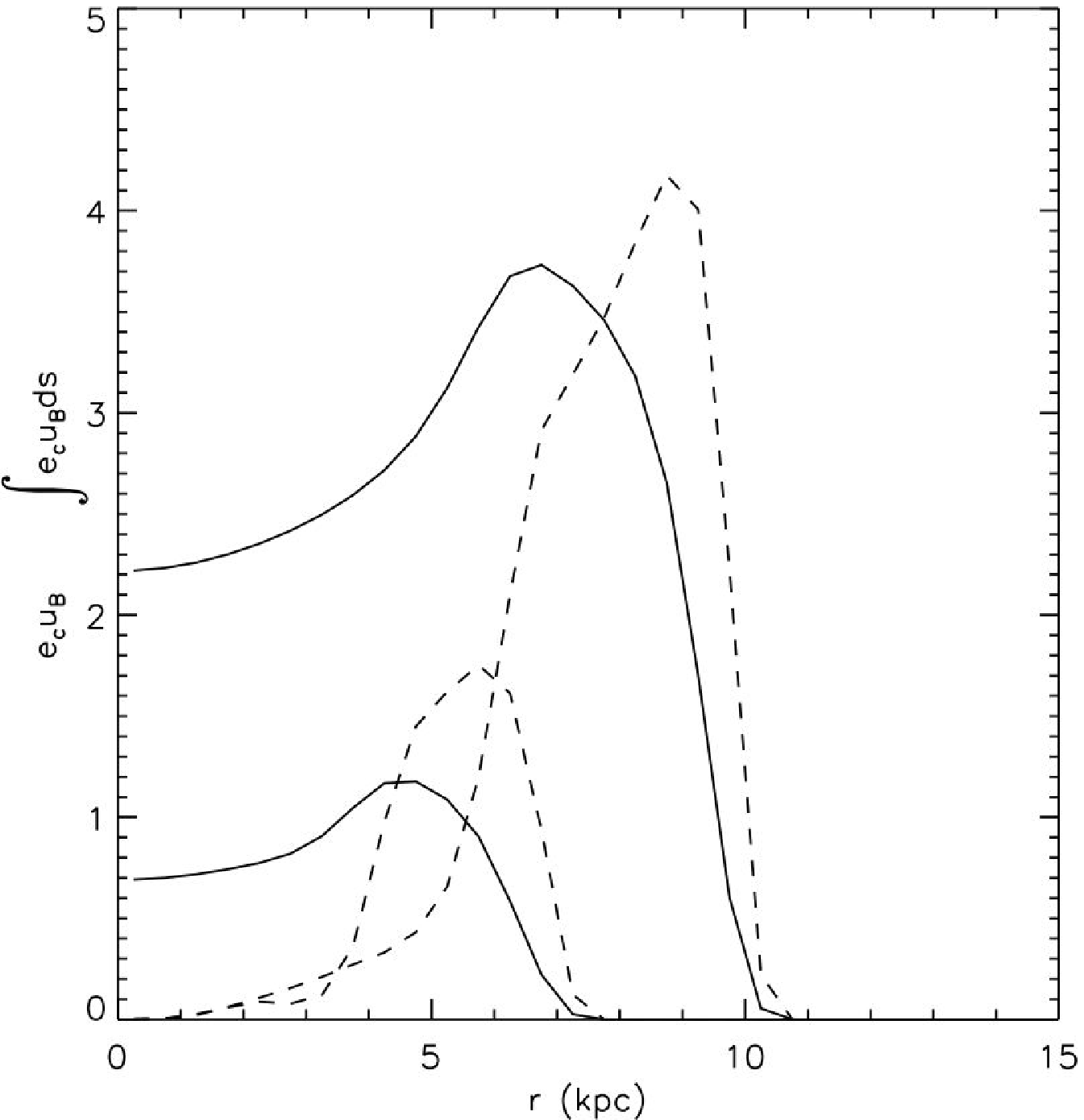}
\vskip.2in
\includegraphics[width=4.in,scale=0.45,angle=0]
{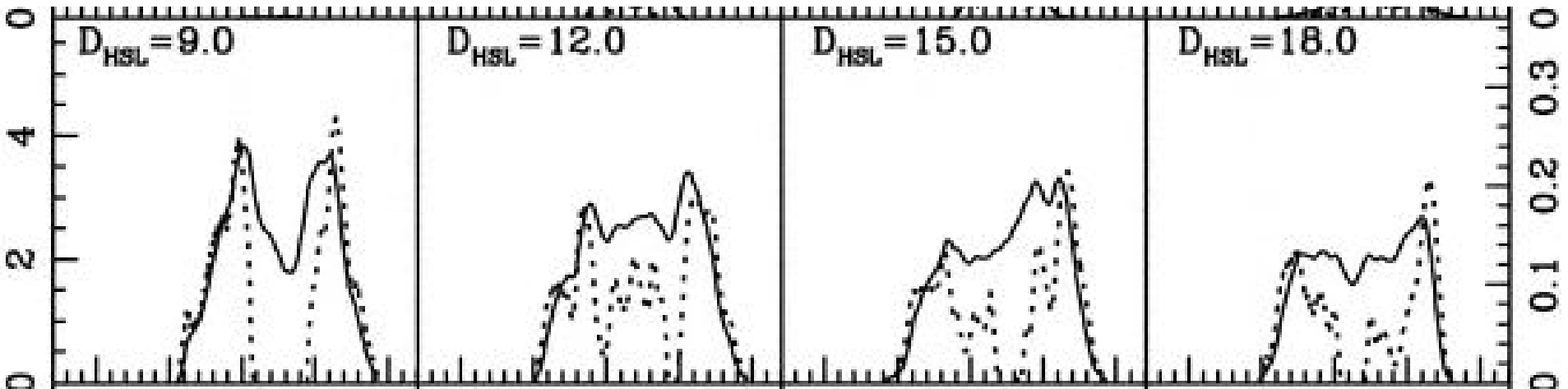}
\vskip.2in
\caption{
{\it Upper panel:} Transverse contours of surface brightness of radio
synchrotron emissivity represented  by $e_cu_B$ ({\it dashed lines})
and corresponding surface brightness 
integrated along a perpendicular line of sight 
$\int e_cu_Bds$ ({\it solid lines}). 
Upper profiles are at $z = 35$ kpc 
and lower profiles 
at $z = 45$ kpc. Units are arbitrary.
{\it Lower panel:} Four typical radio synchrotron surface brightness
contours perpendicular to the Cygnus A Western lobe
$D_{HSL}$ kpc from the hotspot (solid curves) and the corresponding
deconvolved volume emissivity (dotted curves) (Carvalho et
al. 2005). The small horizontal divisions indicate 2 arcseconds 
and the left and right vertical axes are Jy/Beam and
(Jy/beam)/arcsecond respectively.
}
\end{figure}

\clearpage

\begin{figure}
\centering
\includegraphics[width=4.in,scale=0.45,angle=0]
{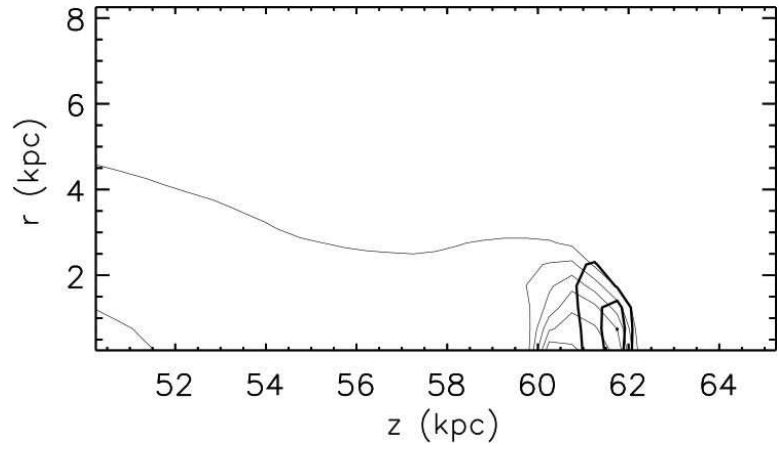}
\caption{Projected offset of computed 
radio and X-ray hotspot structure.
Contours show approximate distribution of 
the IC-CMB X-ray surface brightness $\int e_c ds$ 
({\it light lines}) and 
radio synchrotron surface brightness 
$\int e_cu_B ds$ ({\it heavy lines}).
}
\end{figure}

\clearpage 

\begin{figure}
\centering
\includegraphics[width=4.in,scale=0.45,angle=270]
{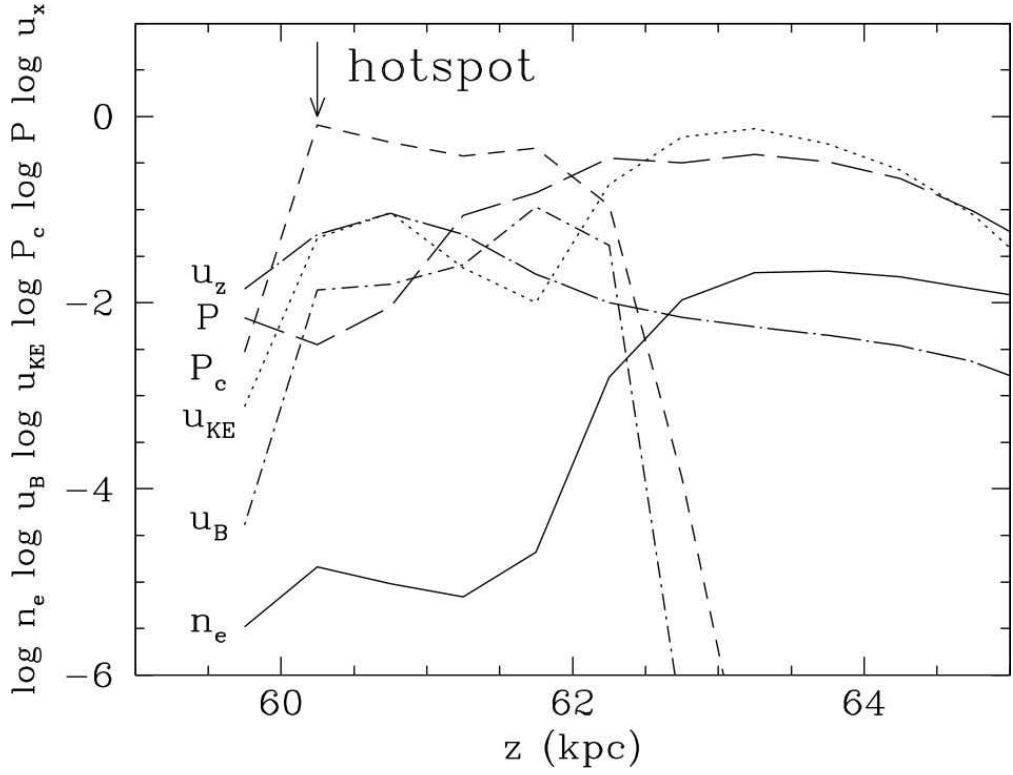}
\caption{Detailed profiles of the hotspot structure 
along the jet (symmetry) axis. 
The gas pressure $P$, CR pressure $P_c$, 
magnetic energy density $u_B = B^2/8\pi$, 
and kinetic energy density $u_{KE} = \rho {(u_z)^2}/2$
are in cgs units increased by a factor of $10^8$. 
The gas velocity $u_z(z)$ in cgs units 
is reduced by a factor $10^{-11}$.
}
\end{figure}

\clearpage

\begin{figure}
\centering
\includegraphics[width=4.in,scale=0.45,angle=0]
{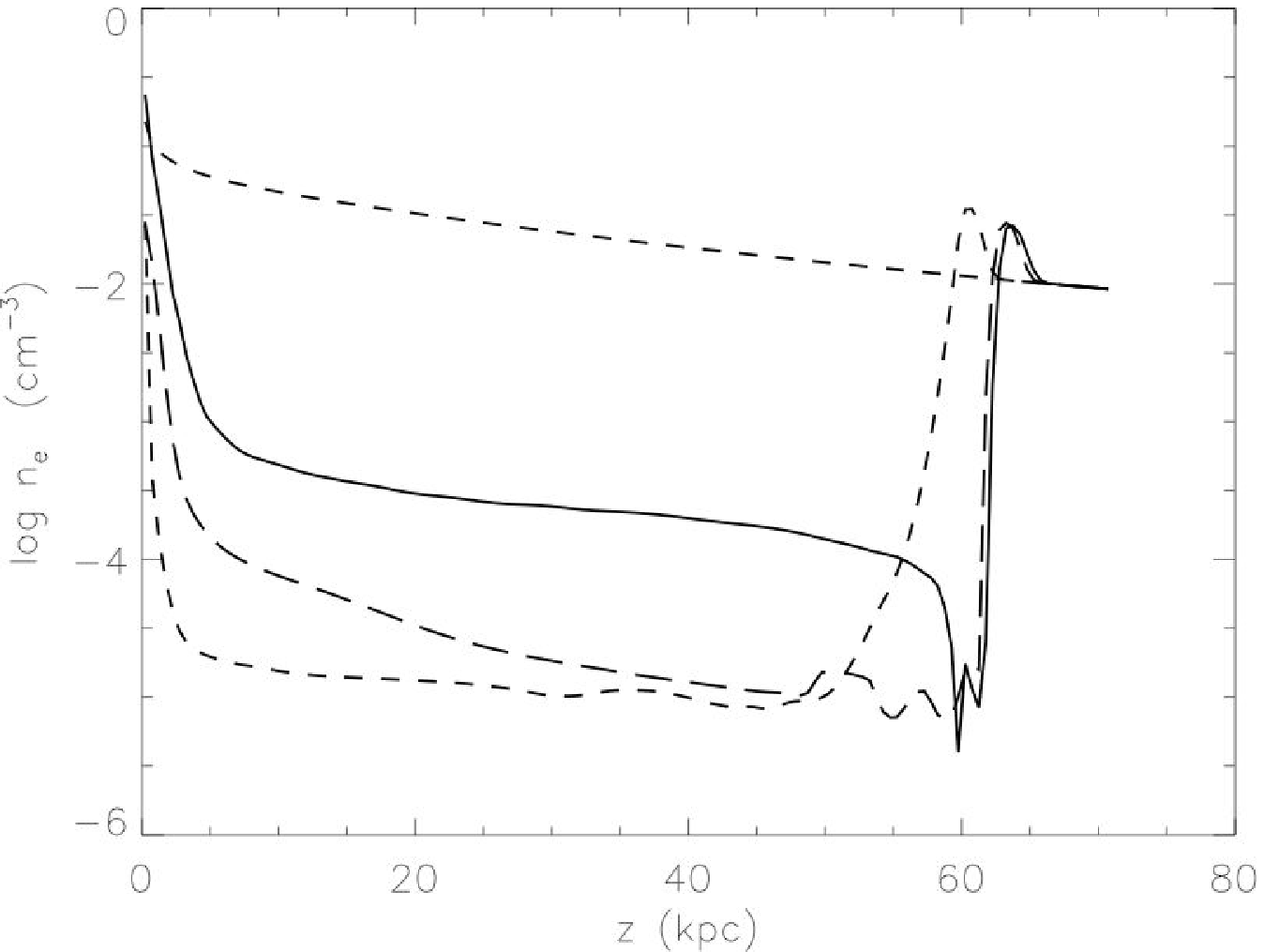}
\includegraphics[width=3.0in,scale=0.45,angle=270]
{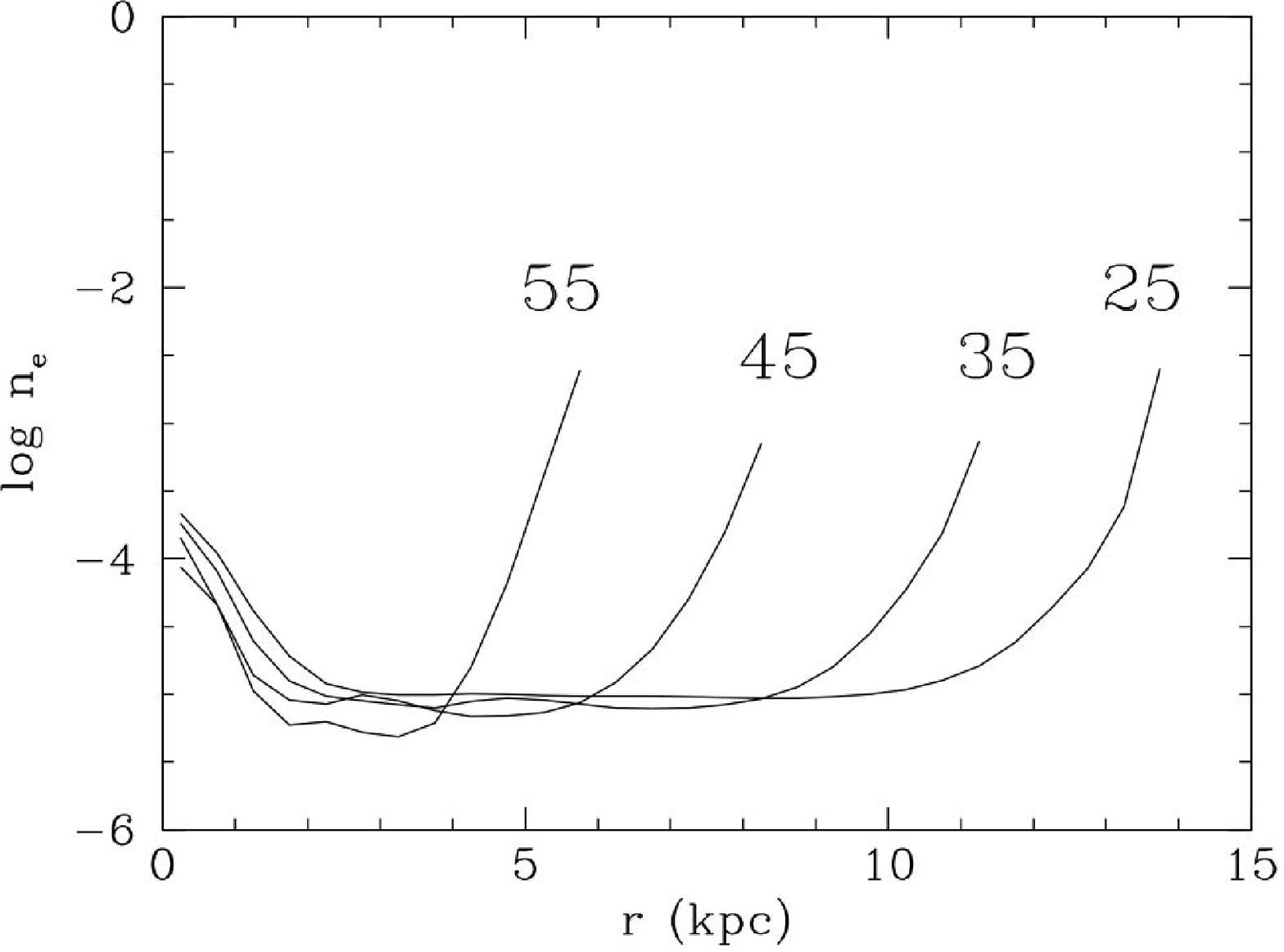}
\caption{{\it Upper panel:} 
Gas density profiles parallel to jet axis
of initial cluster atmosphere ({\it upper short dashed line});
at 10 Myrs along the jet axis ({\it solid line});
at parallel cut at $r = 1.75$ kpc ({\it long dashed line});
at parallel cut at $r = 4.75$ kpc ({\it lower short dashed line}).
{\it Lower panel:} 
Gas density profiles perpendicular to the jet axis at 
$z = 55$, 45, 35 and 25 kpc. 
Outer density profiles terminate at the edge of the X-ray cavity 
defined as $r$ where $n_e(z,r)$ is one third of the 
gas density in the original cluster atmosphere.
}
\end{figure}

\end{document}